\documentclass{aa}  

\usepackage{xcolor}

\usepackage{graphicx}
\usepackage{txfonts}
\usepackage{natbib}
\bibpunct{(}{)}{;}{a}{}{,}

\defcitealias{Hathaway2011}{H11}
\defcitealias{Cloutier2023}{Paper~I}
\defcitealias{Cloutier2024}{Paper~II}

\begin{document}

\title{The rush to the poles\\ and the role of magnetic buoyancy in the solar dynamo}
\titlerunning{The rush to the poles and the role of magnetic buoyancy in the solar dynamo}
\authorrunning{S. Cloutier et al.}

\author{S. Cloutier$^1$,
          R. H. Cameron$^1$,
          \and
          L. Gizon$^{1,2}$
          }

\institute{Max-Planck-Institut f{\"u}r Sonnensystemforschung, Justus-von-Liebig-Weg 3, D-37077 G{\"o}ttingen, Germany\\
              \email{cloutier@mps.mpg.de} 
              \and
              Institut f\"ur Astrophysik und Geophysik, Georg-August-Universit\"at G\"ottingen, D-37077 G{\"o}ttingen, Germany}

\abstract
{The butterfly diagram of the solar cycle exhibits a poleward migration of the diffuse magnetic field resulting from the decay of trailing sunspots. It is one component of what is sometimes referred to as the "rush to the poles" and is responsible for the reversal and build up of the polar cap fields.}
{We investigate under which conditions the rush to the poles can be reproduced in flux-transport Babcock-Leighton dynamo models.  We also consider other observational consequences of the different mechanisms  for reproducing the rush to the poles.}
{We identify three main ways to achieve the rush to the poles: a flux emergence probability that decreases rapidly with latitude; a threshold in subsurface toroidal field strength below which the toroidal flux emerges only slowly and above which the emergence rate is high; and an emergence rate which depends on the mean magnetic field squared, mimicing magnetic buoyancy. We implement these three mechanisms in a 2D Babcock-Leighton flux transport dynamo model incorporating toroidal flux loss and deep downward turbulent pumping. Moreover, we directly compare the observational sunspot zone migration law with what our models predict.}
{We find that all three mechanisms lead to solar-like butterfly diagrams, but which present notable differences between them. The shape of the butterfly diagram is very sensitive to model parameters for the threshold prescription, while most models incorporating magnetic buoyancy converge to very similar butterfly diagrams, with butterfly wings widths of $\lesssim\pm 30^\circ$, in very good agreement with observations. With turbulent diffusivities above $35~\text{km}^2/\text{s}$ but below about $40~\text{km}^2/\text{s}$, buoyancy models are strikingly solar-like. The threshold and magnetic buoyancy prescriptions make the models non-linear and as such can saturate the dynamo through latitudinal quenching -- where emergences at higher latitudes are less efficient at transporting field across the equator and hence less efficient in reversing the polar fields -- although only the latter can do so when emergence loss is turned off. The period of the models involving buoyancy is independent of the source term amplitude, but emergence loss increases it by $\simeq 60\%$. The models, with the right advection amplitude and turbulent diffusivity, match very well the observational equatorward migration law.} 
{For the rush to the poles to be visible, a mechanism suppressing (enhancing) emergences at high (low) latitudes must operate. It is not sufficient that the toroidal field be stored at low latitudes for emergences to be limited to low latitudes. Magnetic buoyancy appears to be the most promising non-linearity as models incorporating it produce the most solar-like butterfly diagrams, with the exact width of the butterly wings being roughly independent of model parameters. Dynamo saturation is achieved by a competition between latitudinal quenching and a quenching due to the tilt of the mean bipolar magnetic region. From these models we infer that the Sun is not in the advection-dominated regime, but also not in the diffusion-dominated regime. The cycle period is set through a balance between advection, diffusion and flux emergence, in a way that agrees with the observational sunspot zone migration law. The latter seems to imply that the toroidal field is indeed stored in the equatorial region of the lower convection zone.}

               
\maketitle

\section{Introduction} \label{sect:intro}

The solar cycle is understood as being driven by a self-exciting fluid dynamo located somewhere inside the convection zone of the Sun \citep[e.g.][]{Charbonneau2014}. \citet[][hereafter \citetalias{Cloutier2023}]{Cloutier2023} have built a 2D Babcock-Leighton (BL) flux-transport dynamo (FTD) model that could self-consistently produce relatively narrow butterfly wings, without imposing a preference for emergences to take place at the observed low latitudes. This was achieved through turbulent pumping reaching deep down to the location where the meridional flow changes direction. However, they found that this linear BL FTD model lacked the so-called "rush to the poles" \citep{Ananthakrishnan1954,Altrock1997}, which we, in the dynamo context, define as the poleward migration of the diffuse magnetic field resulting from the decay of the trailing sunspots.

Surface flux transport (SFT) models \citep{Yeates2023} reproduce well the rush to the poles. In these models, observed or modeled active regions are deposited on the surface where they are passively transported by advection and diffusion. The rush to the poles is hence a consequence of the properties of observed emergences. In FTD models, the properties of emergences are a property of the model, and depend on how the emergence process is parametrized. As such, and as shown in \citetalias{Cloutier2023}, they do not necessarily reproduce the rush to the poles.

The questions we address in this paper are which properties of the observed emergences are necessary to reproduce the rush to the poles and what are the constraints it places on the dynamo process (particularly the conversion of the toroidal to poloidal flux). We identify three possible mechanisms by which the rush to the poles can be achieved: a latitudinal sunspot emergence probability caused by stability of the toroidal field at mid- to high latitudes \citep[e.g.][and references therein; see also \citealt{Kitchatinov2020}]{KC2016}, a threshold in subsurface toroidal field strength between a slow and a fast regime of flux emergence \citep{CS2020,Biswas2022}, and a proportionality of the emergence rate to a power of the ratio of the toroidal to equipartition magnetic field strengths, the latter mechanism being motivated by magnetic buoyancy \citep{Stix1972,Parker1975,UR1976}.

Lastly, we note that the location of maximal toroidal flux density is a good proxy for the central latitude of the sunspot belt, so that the equatorward drift of the former can therefore be compared to the observed drift of the latter. This migration has been shown by \citet{Waldmeier1939,Waldmeier1955} to be universal regardless of cycle strength, the functional form of which was determined by \citet{Hathaway2011}.

\section{Model}
\subsection{Dynamo equations}
The model we use is the same as in \citetalias{Cloutier2023}. The equations we solve are the 2D axisymmetric  mean-field dynamo equations for the $\phi$-component of the poloidal vector potential $A$ and the toroidal component $B$ of the large-scale magnetic field:
\begin{equation}
\frac{\partial A}{\partial t}=-\frac{\boldsymbol{u}_p}{\varpi}\cdot\nabla (\varpi A)+\eta\left(\nabla^2-\frac{1}{\varpi^2}\right)A+S, \label{Aeq}
\end{equation}
\begin{equation}
\begin{aligned}
\frac{\partial B}{\partial t}=&-\varpi\boldsymbol{u}_p\cdot\nabla\left(\frac{B}{\varpi}\right)+\eta\left(\nabla^2-\frac{1}{\varpi^2}\right)B+\frac{1}{\varpi}\frac{\partial (\varpi B)}{\partial r}\frac{\mathrm{d}\eta}{\mathrm{d}r}\\&-B\nabla\cdot\boldsymbol{u}_p+\varpi [\nabla\times (A\boldsymbol{\hat{e}}_{\phi})]\cdot\nabla\Omega -L, \label{Beq}
\end{aligned}
\end{equation}
where $\varpi =r\sin{\theta}$. 
The effective meridional velocity $\boldsymbol{u}_p=\boldsymbol{u}_m+\boldsymbol{\gamma}$ is the sum of the meridional flow $\boldsymbol{u}_m$ and  turbulent pumping $\boldsymbol{\gamma}$, $\Omega$ is the local rotation rate (including differential rotation), $\eta$ is the turbulent diffusivity, and $S$ and $L$ are the BL source and loss terms, respectively. A feature of our model is the inclusion of this toroidal field loss term; it is associated with the emergence of active regions, which give rise to the BL mechanism. Such a loss term was taken into account in the original model of \citet{Leighton1969}, but it was quickly abandoned as it was deemed not to be of significance to the magnetic budget of the Sun. However, this assumption was recently shown to be wrong by \citet{CS2020}, as they determined the corresponding timescale to be commensurate with the 11-year solar cycle. This observational result was further found to be naturally reproduced with the linear loss term of \citetalias{Cloutier2023}.

\subsection{Differential rotation and meridional circulation}
\begin{figure}
\resizebox{\hsize}{!}{\includegraphics{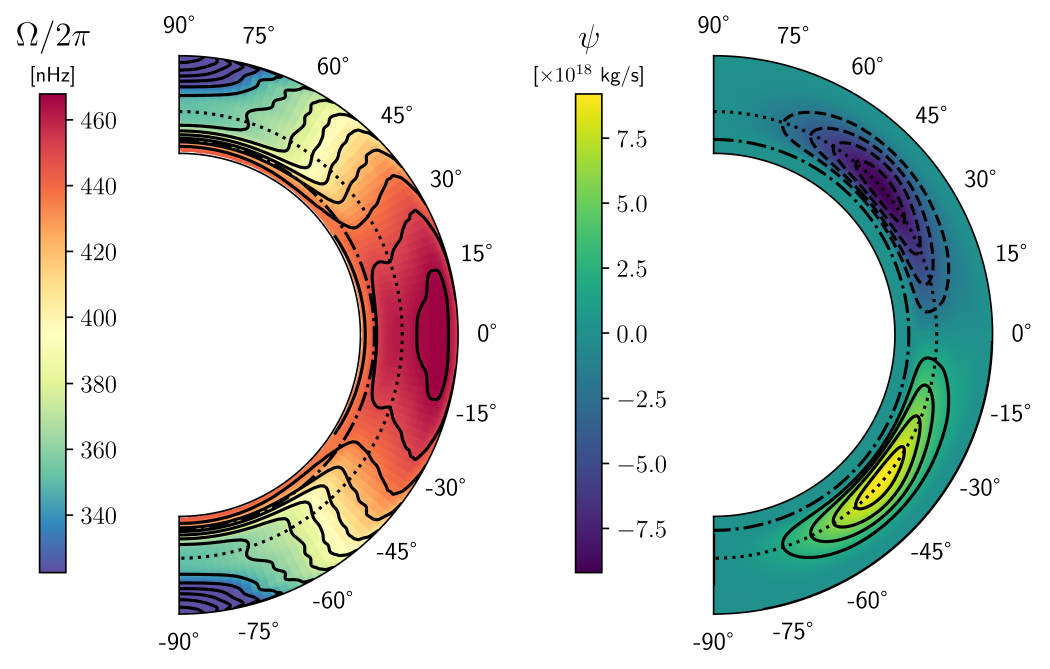}}
\caption{Rotation profile of \citet{LS2018} obtained from HMI data (left) and cycle-averaged and symmetrized stream function of the helioseismic meridional flow inversions of \citet[][right]{Gizon2020}. For the latter, positive values represent clockwise circulation and negative anticlockwise. The dash-dotted and dotted lines represent the approximate locations of the tachocline at $0.7R_{\odot}$ and reversal of the meridional flow direction at $0.8R_{\odot}$, respectively.}
\label{fig:flows}
\end{figure}

For the differential rotation profile we here use the helioseismic measurement of \citet{LS2018} obtained from HMI data, which is shown in the left panel of Fig. \ref{fig:flows}. We do not apply any sort of mask to the profile to "correct" for the high latitudes \citep[e.g.][]{MJ2009}, although the exact rotation rate at these latitudes is rather important (cf. \citetalias{Cloutier2023}).

The meridional flow profile we use is the same as in \citetalias{Cloutier2023} and is shown in the right panel of Fig. \ref{fig:flows}. It is constructed from the helioseismic inversions of \citet{Gizon2020} by symmetrizing the profiles of cycles 23 and 24 across the equator and averaging them. The dotted line represents the location where the meridional flow changes directions at a radius of about 0.8$R_{\odot}$. The toroidal field is in our models essentially stored below that depth as our turbulent pumping profile reaches down to and stops at that depth (see the next section and \citetalias{Cloutier2023} for a discussion).

\subsection{Parameterization of turbulent effects} \label{sect:turb}
The turbulent parameterizations are the same as in \citetalias{Cloutier2023} (see also references therein). 
The turbulent diffusivity profile is expressed as
\begin{equation}
\begin{aligned}
\eta (r)=\eta_{\text{RZ}}+&\frac{\eta_{\text{CZ}}-\eta_{\text{RZ}}}{2}\left[1+\text{erf}\left(\frac{r-0.72R_\odot}{0.012R_\odot}\right)\right]\\
+&\frac{\eta_{R_\odot}-\eta_{\text{CZ}}-\eta_{\text{RZ}}}{2}\left[1+\text{erf}\left(\frac{r-0.95R_\odot}{0.01R_\odot}\right)\right],
\end{aligned}
\end{equation}
where $\eta_{\text{RZ}}=0.1$ km$^2$/s, $\eta_{R_\odot}=350$ km$^2$/s and $\eta_{\text{CZ}}$ are respectively the radiative core, surface, and bulk values of the turbulent diffusivity. It was found in \citetalias{Cloutier2023} that, in this class of models, bulk diffusivities significantly larger than $10~$km$^2$/s are not possible because the large radial shear of the observed deep meridional flow gives rise to an effective diffusivity as high as $\simeq 150~\text{km}^2/\text{s}$. For most of the models presented in this paper, the value of $\eta_{\text{CZ}}$ is thus fixed at $10~$km$^2$/s. It will however be shown that the bulk diffusivity can be significantly increased in the non-linear models, particularly those invoking magnetic buoyancy.

For turbulent pumping we again use the following single step profile:
\begin{equation} \label{eq:pump}
\boldsymbol{\gamma}=-\frac{\gamma_0}{2}\left[1+\text{erf}\left(\frac{r-r_{\gamma}}{0.01R_\odot}\right)\right]\boldsymbol{\hat{e}}_r,
\end{equation}
where $r_{\gamma}=0.785R_{\odot}$ corresponds to the depth at mid-latitudes where the meridional flow changes direction.

\subsection{BL source and loss terms}
The BL source and loss terms are given by:
\begin{align}
S(r,\theta,t)=~&
     f_r^S(r)\sin^n\theta\sin\delta\frac{b(\theta,t)/R_{\odot}}{\tau_0}, \label{eq:s}\\
L(r,\theta,t)=~&f_r^L(r) \sin^n\theta\cos\delta
\frac{B(r,\theta,t)}{\tau_0}, \label{eq:l}
\end{align}
where:
\begin{align}
f_r^S(r)=~&\frac{1}{2}\left[1+\text{erf}\left(\frac{r-0.85R_\odot}{0.01R_\odot}\right)\right],\\
f_r^L(r)=~&\frac{1}{2}\left[1+\text{erf}\left(\frac{r-0.70R_\odot}{0.01R_\odot}\right)\right],
\end{align}
and $b$ is the toroidal flux density inside the convection zone:
\begin{equation}
b(\theta,t)=\int_{0.7R_\odot}^{R_\odot} B(r,\theta,t) r \mathrm{d}r.
\end{equation}
Except for the exponent $n$ in the $\sin^n\theta$ terms, these expressions are the same as in \citetalias{Cloutier2023} and their derivation can be found there. 

The source term above (S) is that of the $\phi$-component of the poloidal vector potential $A$. But the more physically relevant quantity is the \textit{radial field} generation rate at the surface \citep[][hereafter \citetalias{Cloutier2024}, see also Appendix \ref{sect:app}]{Cloutier2024}, which from the definition of $A$ (Eq. (1) of \citetalias{Cloutier2023}) is readily found to be given by
\begin{equation}
    S_r(R_{\odot},\theta,t)=\frac{1}{R_{\odot}^2\sin\theta}\frac{\partial}{\partial\theta}\left(\sin^{n+1}\theta\sin\delta\frac{b(\theta,t)}{\tau_0}\right). \label{eq:sr}
\end{equation}
Regularity of the source term at the poles is ensured as long as $n\geq 1$ (see second form of Eq. (\ref{eq:sra})).

\subsubsection{Latitudinal emergence probability}
The  $\sin^n\theta$ terms allow us to study cases where emergence at low latitudes is explicitly imposed \citep{KC2016}. In the first set of models, we will vary the value of $n$ from 1 to 12. The resulting models are  linear, and we will choose the values of $\tau_0$ and $\gamma_0$ so that they are critical with a period of 12 years (the average period of cycles 23 and 24). The linearity of the model also means that the fields can be arbitrarily scaled. We normalize the fields so that the maximum net toroidal flux in one hemisphere (here the northern) is $\text{max}(\Phi)=5\times 10^{23}$ Mx, where
\begin{equation}
\Phi(t)=\int_0^{\pi}b(\theta,t)\mathrm{d}\theta.
\end{equation}
This value is consistent with the estimates of \citet{CS2015}.

\subsubsection{Two-regime threshold} \label{sect:modtworeg}
In the second set of models the value of $n$ will be fixed to 1, representing an emergence probability constant per unit length of toroidal field lines. But we will set a threshold $B_{\text{thresh}}$ in the average toroidal field $\overline{B}$ between two emergence regimes $\tau_0^{\text{slow}}$ and $\tau_0^{\text{fast}}$. This second emergence rate model is motivated by the observed toroidal field maps which show much stronger values of surface $B_\phi$ when active regions begin to emerge, suggestive of a switch between slow and rapid emergence once some threshold is met \citep[see also][]{CJ2019,Biswas2022}.

We define the average toroidal field for this purpose to be
\begin{equation}
    \overline{B}(\theta,t)=\int_{0.7R_{\odot}}^{R_{\odot}}B(r,\theta,t)r\mathrm{d}r\bigg/\int_{0.7R_{\odot}}^{0.8R_{\odot}}r\mathrm{d}r=\frac{b(\theta,t)}{0.075R_{\odot}^2},
\end{equation}
where we made the approximation that the toroidal field is stored uniformly in the lower half of the convection zone. Then, we introduce a threshold on $\overline{B}$, which is equivalent to one on $b$,
\begin{equation}    
b_{\text{thresh}}=0.075R_{\odot}^2 \overline{B}_{\text{thresh}},
\end{equation}
and use it to determine whether flux emergence should be fast or slow:
\begin{equation}
    \tau_0(\theta,t)=
    \begin{cases}
    \tau_0^{\text{slow}} & b(\theta,t)<b_{\text{thresh}}, \\
    \tau_0^{\text{fast}} & b(\theta,t)\geq b_{\text{thresh}}.
    \end{cases}
\end{equation}
This threshold prescription makes the model non-linear.

\subsubsection{Magnetic buoyancy} \label{sect:buoy}

\citet{Parker1955} and \citet{Jensen1955} independently showed that magnetic flux tubes initially in thermal equilibrium are buoyant. When they are solely resisted by aerodynamic drag, such flux tubes will float to the surface with terminal rise velocities of the order of the Alfvén velocity \citep{Parker1975},
\begin{equation} \label{eq:vb1}
    v_B\sim v_\text{A}=\left(\frac{v_\text{A}}{v_c}\right)v_c=\left|\frac{B}{B_\text{eq}}\right|v_c,
\end{equation}
$v_{\text{A}}=|B|/\sqrt{\mu_0\rho}$ and $v_c$ being respectively the Alfvén velocity and convective velocity given by mixing-length theory \citep[MLT,][]{Prandtl1925,Vitense1953,BV1958}, and $B_{\text{eq}}=\sqrt{\mu_0\rho}v_c$ the equipartition field strength. Taking into account viscous drag and assuming the turbulent viscosity to be given by MLT, the terminal rise velocity of the flux tube is instead of the order of \citep{UR1976}
\begin{equation}
    v_B\sim\left(\frac{v_\text{A}}{v_c}\right)v_\text{A}=\left(\frac{v_\text{A}}{v_c}\right)^2v_c=\left(\frac{B}{B_\text{eq}}\right)^2v_c. \label{eq:vb2}
\end{equation}
Note that for both cases, the relevant magnetic field $B$ is that of the \textit{total} field.

\citet{KP1993}, however, argued that one should instead consider magnetic buoyancy within the framework of mean-field electrodynamics \citep{Moffatt1978,KR1980}. Employing the second-order correlation approximation, the authors found the same buoyant velocity dependence on $(\overline{B}/B_{\text{eq}})^2$ as that found by \citet{UR1976}, but with a quenching term due to magnetic tension. For strong mean toroidal fields the rise velocity decreases as $(\overline{B}/B_{\text{eq}})^{-1/2}$. In taking into account magnetic buoyancy in our model, we will consider both rise velocities given by Eqs. (\ref{eq:vb1}) and (\ref{eq:vb2}). For simplicity we do not consider any effect due to magnetic tension.

\begin{figure*}
\centering
   \includegraphics[width=17cm]{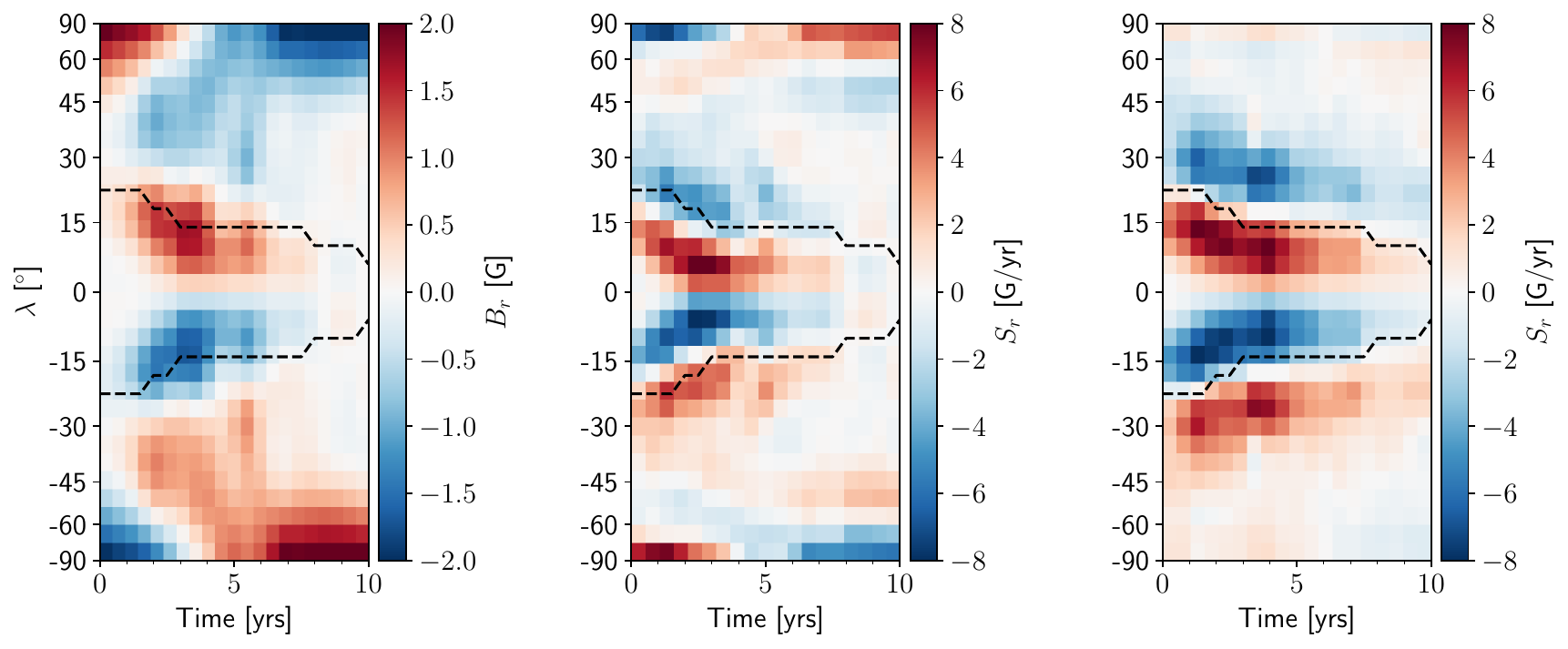}
     \caption{\textit{Left panel:} cycle-averaged butterfly diagram of \citetalias{Cloutier2024}. \textit{Middle panel:} surface radial source term obtained from the left panel by "inverting" a 1D surface flux transport model. \textit{Right panel:} surface radial source term obtained from the surface toroidal field (not shown) using an emergence model. See \citetalias{Cloutier2024} for more information. The dashed lines represent the location of maximum surface toroidal field.}
     \label{fig:sr_obs_hath}
\end{figure*}

Given the implicit condition behind our source and loss terms, that there is a continuous emergence of sunspots over timescales over which the configuration of the mean field changes appreciably (cf. \citetalias{Cloutier2023}), the emergence \textit{rate} should be proportional to a representative buoyant terminal rise velocity inside the convection zone. Hence, under these assumptions, the timescale parameter is given by
\begin{equation}
\tau_0^{-1}=\left|\frac{\overline{B}}{B_\text{eq}}\right|^m\left(\tau_0^b\right)^{-1}=\left|\frac{b}{b_\text{eq}}\right|^m\left(\tau_0^b\right)^{-1},
\end{equation}
where $\tau_0^b$ is now the free parameter. The above prescriptions of \citet{Parker1975} and \citet{UR1976} are given by $m=1$ and $m=2$ respectively. For $m>0$ the source and loss terms become non-linear in the magnetic field and thus can, in \mbox{principle}, \mbox{provide} a saturation mechanism for the dynamo. Because we \mbox{expect} the toroidal field to be concentrated at low latitudes, this prescription should not be too dissimilar to the two-regime threshold. 

This idea has, in essence, first been proposed by \citet{Stix1972} in the context of a non-linear \textit{turbulent} $\alpha\Omega$ dynamo model. Non-linear toroidal field loss due to magnetic buoyancy is not a new feature of $\alpha\Omega$ dynamo models \citep[e.g.][]{SS1989,Moss1990a,Moss1990b,JW1991}. However, our BL source and loss terms are linked (cf. \citetalias{Cloutier2023}), and so magnetic buoyancy must be taken into account in the \textit{source} term as well. \citet{Jouve2010} and \citet{Fournier2018} have explored the effect of an emergence \textit{delay} dependent on the magnetic energy of the toroidal field at the bottom of the convection zone, while \citet{KM2017} considered a varying emergence rate, but for 3D models already including tilt quenching and neglecting emergence loss.

\section{Observational constraints}
\subsection{Sunspot number proxy}
Flux emergence both removes toroidal magnetic flux from the solar interior and creates poloidal field (and sunspots) at the solar surface. The number of sunspots formed at the surface can be estimated from the amount of toroidal magnetic flux which is lost in the process. To do so, we first note that typical active regions sizes are around $d_{\text{AR}}=100$~Mm. Hence, the rate at which flux is generated in active regions is 
\begin{equation}
\frac{\mathrm{d}\Phi_{\text{AR}}}{\mathrm{d}t}(t)=\frac{2\pi R_\odot}{d_{\text{AR}}}\int_0^{\pi}\int_{0.70 R_{\odot}}^{R_{\odot}}|L(r,\theta,t)|r\mathrm{d}r\mathrm{d}\theta,
\end{equation}
where $L$ is the BL loss term defined by Eq. (\ref{eq:l}). We then divide this quantity by a representative value of the flux contained in a sunspot, which we take to be $\Phi_\text{S}=10^{21}~\text{Mx}$, so that we have a number of sunspots per year being generated. Assuming these sunspots to have a lifetime of one month yields our sunspot number proxy:
\begin{equation}
R=\frac{1}{\Phi_\text{S}}\int_{t-1~\mathrm{mth}}^{t}\frac{\mathrm{d}\Phi_{\text{AR}}}{\mathrm{d}t}(t^\prime)\mathrm{d}t^\prime.
\end{equation}

A quantity we will consider is the ratio of the sunspot number at activity minimum to maximum,
\begin{equation}
    \Delta R=\frac{\text{min}(R)}{\text{max}(R)}.
\end{equation}
It is a measure of the overlap between cycles. 

\subsection{Equatorward drift of the sunspot zones}
An important discovery is that of \citet{Waldmeier1939, Waldmeier1955}, who found that the equatorward migration of the central heliographic latitude of the sunspot zones, $\lambda_\text{c}$, is alike for all cycles, regardless of cycle strength of phase. Indeed, choosing the \mbox{reference} times of individual cycles to be the times of the first appearance of sunspots belonging to those particular cycles \citep{Waldmeier1935}, rather than the times of activity minimum or maximum, the $\lambda_\text{c}$-curves superpose; the equatorward drift of the activity belts follows a standard path. This feature was further confirmed by \citet[][hereafter \citetalias{Hathaway2011}]{Hathaway2011}. Employing a \mbox{parametric} function devised by \citet{Hathaway1994} to fit the monthly sunspot number of individual cycles, \citetalias{Hathaway2011} could determine the start times of cycles 12 to 23 and rediscover the old finding of \citet{Waldmeier1939,Waldmeier1955}. Moreover, \citetalias{Hathaway2011} could fit the centroid curves as
\begin{equation}
    \lambda_\text{c}(t)=28^\circ\exp\left(-\frac{t-t_0}{90~\mathrm{mths}}\right), \label{eq:hlaw}
\end{equation}
where time is defined in months and the centroid latitude is $\lambda_\text{c}$.

This standard equatorward migration law is an observational constraint on dynamo models. For our models, we will assume the centroid latitude of the sunspot zone $\lambda_\text{c}$ to be given by the latitude where the toroidal flux density $b$ is maximum. This latitude closely corresponds to the emergence location of the mean bipolar magnetic region (BMR), as demonstrated in Appendix \ref{sect:app} and as inferred from the observations (\citetalias{Cloutier2024}, see Fig. \ref{fig:sr_obs_hath}). Since it is difficult to define exactly when a cycle begins (the functional form for the sunspot number given by \citetalias{Hathaway2011} may not approximate our sunspot number proxy), we choose the value of $t_0$ so as to obtain the best fit between the centroid latitude curves. As in \citetalias{Hathaway2011}, we find that the start times of our model cycles do not coincide with an activity minimum, here defined as the times where our sunspot proxy $R$ is minimum. 

\begin{figure*}
 \centering
 \includegraphics[width=17cm]{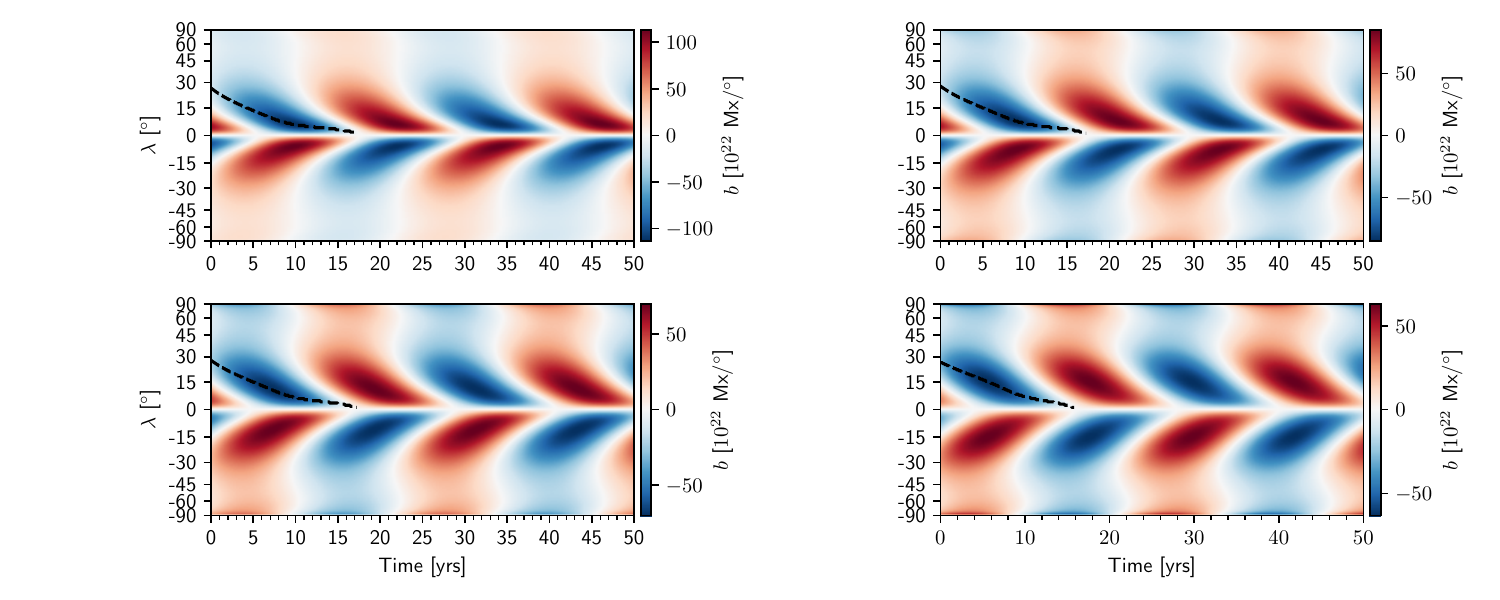}
\caption{Time-latitude diagrams of the toroidal flux density $b$ for different values of $n$: 1 (upper-left panel), 3 (upper-right), 6 (lower-left), and 12 (lower-right).}
\label{fig:nf_b}
\end{figure*}

\begin{figure*}
 \centering
 \includegraphics[width=17cm]{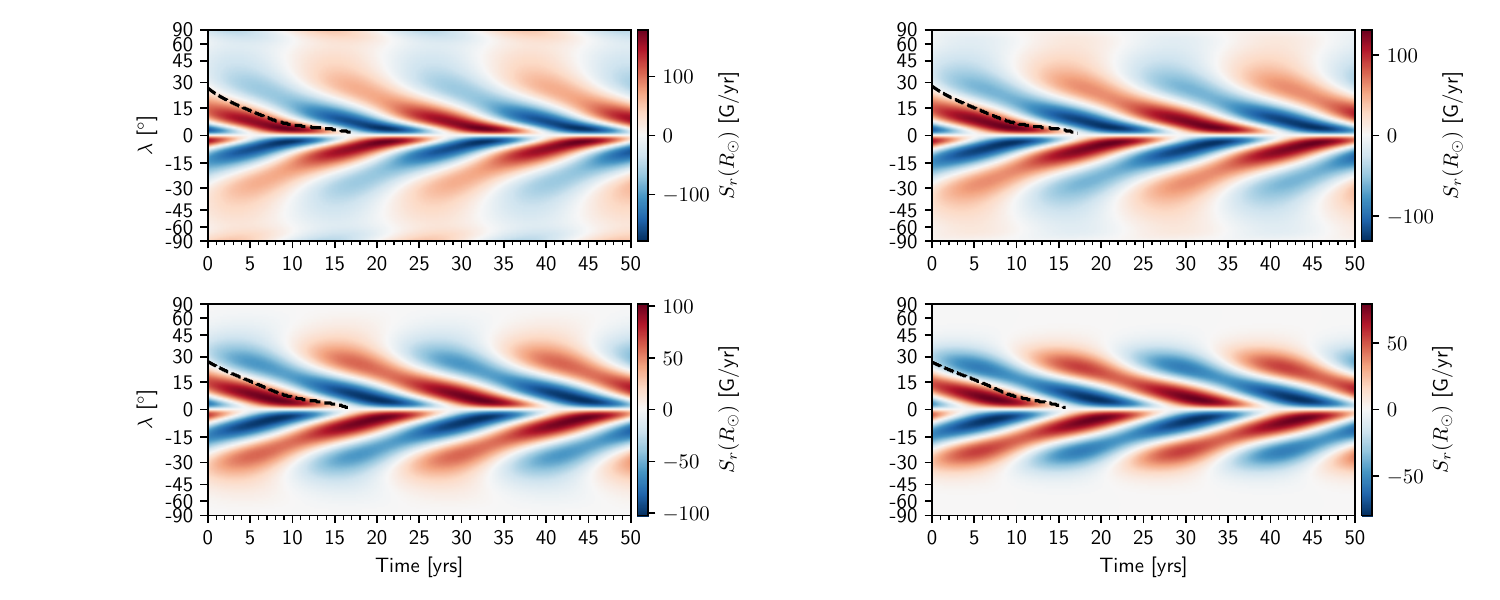}
\caption{Time-latitude diagrams of the surface radial source term $S_r(R_\odot)$ for different values of $n$: 1 (upper-left panel), 3 (upper-right), 6 (lower-left), and 12 (lower-right).}
\label{fig:nf_sr}
\end{figure*}

\begin{figure*}
 \centering
 \includegraphics[width=17cm]{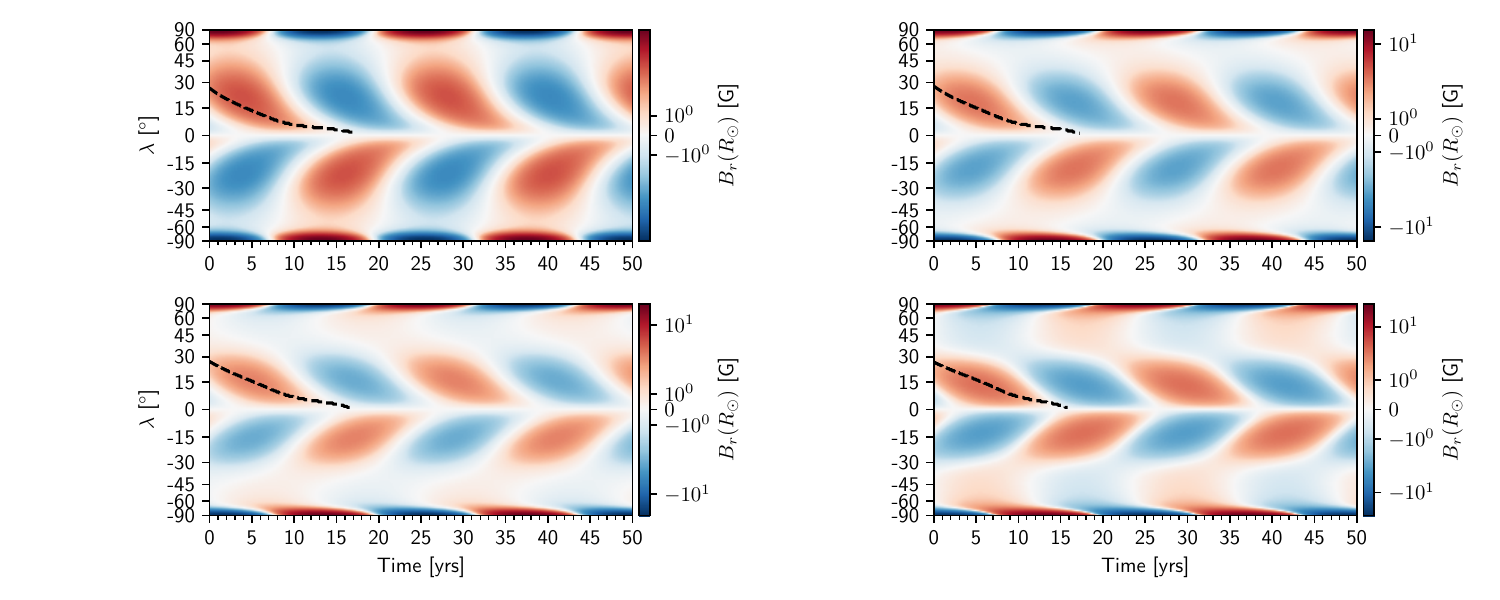}
\caption{Time-latitude diagrams of the surface radial field $B_r(R_\odot)$ for different values of $n$: 1 (upper-left panel), 3 (upper-right), 6 (lower-left), and 12 (lower-right). The color scale is logarithmic to better show the different features.}
\label{fig:nf_butt}
\end{figure*}

\subsection{Polar and butterfly fields}

Important observational constraints are the field strengths on the solar surface. However, one must be careful to actually compare the same quantities. The existence of an universal law of sunspot belt migration means that cycles can be averaged together in phase. This fact was pointed out in \citetalias{Cloutier2024}, where we further obtained the cycle-averaged butterfly diagram presented in Fig. \ref{fig:sr_obs_hath}. The mean field strengths of the butterfly wings at cycle maximum $B_b$ are around 2~G. Now, because of the open flux problem \citep{Linker2017}, the mean polar field at cycle minima $B_p$ could be anywhere in the range of 3-15~G \citep[][although it is more likely to be on the higher end, if not higher -- on that see \citealt{Sinjan2024}]{Petrie2015}.

\subsection{Toroidal flux loss timescale and cycle phase of polar maxima}
The toroidal flux loss timescale $\tau$ and its contributions due to emergence $\tau_L$ and diffusive $\tau_\eta$ loss through the surface are as defined in \citetalias{Cloutier2023}. We here again take their combination to be $\tau^{-1}=\tau_L^{-1}+\tau_\eta^{-1}$. In this paper we calculate these timescales at the time when the polar field reverses. The phase difference between the polar field and cycle maxima $\Delta\phi$ is also defined in \citetalias{Cloutier2023}.

\section{Results \label{sect:results}}

All solutions we present are predominantly of dipolar parity, although our model does allow different parities. In terms of the rush to the poles, it has only been observed for the dipole mode. Therefore we restrict our analysis and discussion to the dipole dynamo mode.

\subsection{Models with an explicit preference for emergence at low latitudes \label{sect:emerprob}}

To look into the effect of $n$ on the latitudinal emergence probability, we computed critical linear models ($B_\text{thresh}=0$ and $m=0$) with values of $n=1,3,6$ and $12$ (see Eqs. \ref{eq:s} and \ref{eq:l}). The solutions are shown in Figs. \ref{fig:nf_b}, \ref{fig:nf_sr} and \ref{fig:nf_butt}, representing the toroidal flux density $b$, the surface radial source term $S_r$, and the surface radial field $B_r$, respectively. The location of the maximum toroidal field density is represented by a dashed line. 

\begin{figure}
\resizebox{\hsize}{!}{\includegraphics{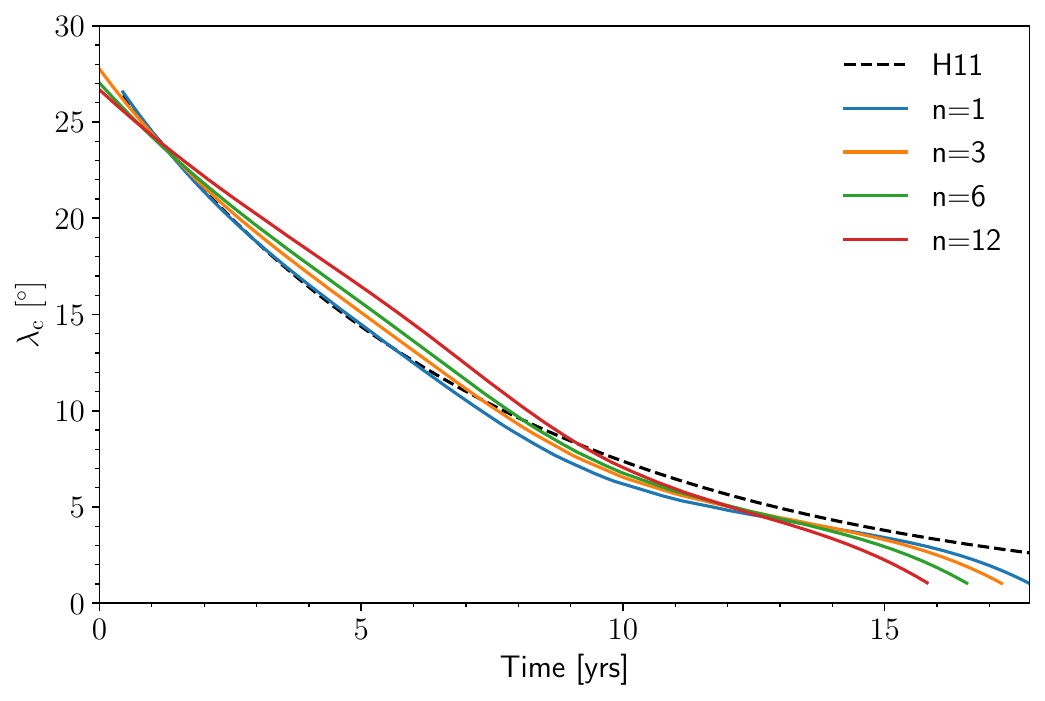}}
\caption{Comparison of the \citetalias{Hathaway2011} standard law of sunspot zone migration with that obtained by linear models.}
\label{fig:hath_lin}
\end{figure} 

The $n=1$ case is essentially the same as the reference model of \citetalias{Cloutier2023}, the only difference being the differential rotation profile. The lack of a distinct "rush to the poles" in the butterfly diagram is obvious. As expected, increasing the value of $n$ forces emergences to occur at increasingly lower latitudes. A high-latitude rush to the poles becomes noticeable at around $60^\circ$ near cycle maximum for the $n=3$ case. By $n=12$, the rush starts at about $30^\circ$. The presence of a rush to the poles is not a necessary consequence of the toroidal field being mostly stored below $30^\circ$. In fact, as it can be appreciated in Fig. \ref{fig:nf_b}, the toroidal field is \textit{more} confined to low latitudes in the $n=1$ case. Especially in the $n=12$ case, there is strong toroidal field at the poles and significant field strengths at mid-latitudes. This is because larger values of $n$ cause more cross-equator poloidal flux cancellation and hence stronger polar fields. The radial shear present throughout the whole depth of the polar convection zone then generates this high-latitude toroidal field. The pumping required to obtain critical 12-year periodic solutions also decreases with increasing $n$, causing more poloidal field to reach the poles. As discussed in \citetalias{Cloutier2023}, stronger pumping is required in the $n=1$ case to concentrate the toroidal field at low latitudes, ensuring dynamo action. This is not necessary when $n>1$ as high latitude emergences are then inhibited.

Fig. \ref{fig:nf_sr} shows the poloidal field generation rate of the models, which must be compared to the observationally-inferred one presented in \citetalias{Cloutier2024} and shown in Fig. \ref{fig:sr_obs_hath}. As in \citetalias{Cloutier2024}, we clearly see two large regions where mean poloidal field of opposite polarities is being generated at each timestep. The lack of a rush to the poles in the $n=1$ case is due to the emergence of flux at all latitudes; even though the toroidal field is concentrated near the equator, there is still enough emergence happening at higher latitudes to prevent the appearance of a rush to the poles. The trailing spot fields migrating polewards from say $30^\circ$ are weak, being spread out over about $60^\circ$ in latitude, and so the much stronger leading spot fields emerging at low latitude emergences dominate over them at mid-latitudes. In other words, what we see at mid-latitudes is the poleward migration of the \textit{leading} polarity field. Increasing $n$ reduces the high-latitude emergence rate and hence allows the trailing polarity flux being advected towards the poles to dominate at mid-latitudes and above.

\begin{table}
\caption{Input parameters of threshold models}             
\label{table:thresh}    
\centering                                     
\begin{tabular}{c c c c c}          
\hline\hline                       
Model & TA & TB & TC & TD \\    
\hline                                 
    $\tau_0^\text{fast}~[\text{yrs}]$ & 10 & 9 & 15 & 5 \\      
    $\gamma_0~[\text{m/s}]$ & 15 & 12.5 & 25 & 12.5 \\
    $\eta_\text{CZ}~[\text{km}^2/\text{s}]$ & 10 & 10 & 20 & 10 \\
    $B_\text{thresh}~[\text{kG}]$ & 1 & 1 & 1 & 1.35  \\
\hline                                            
\end{tabular}
\end{table}

\begin{table}
\caption{Output quantities of threshold models}              
\label{table:thresh_out}      
\centering                                   
\begin{tabular}{c c c c c c c}        
\hline\hline                       
Model & TA & TB & TC & TD \\   
\hline                                   
    $P~[\text{yrs}]$ & 12.1 & 12.1 & 13.1 & 12 \\  
    $B_p~[\text{G}]$ & 22.7 & 18.7 & 19.5 & 10.9 \\
    $B_b~[\text{G}]$ & 4.8 & 4.1 & 3 & 4.7 \\
    $\Phi_\text{m}~[10^{23}~\text{Mx}]$ & 5.8 & 4.3 & 6.2 & 3.9 \\
    $\text{max}(R)$ & 232 & 152 & 184 & 39 \\
    $\text{min}(R)$ & 87 & 39 & 63 & 20 \\
    $\Delta R$ & 0.37 & 0.26 & 0.34 & 0.51 \\
    $\tau_L~[\text{yrs}]$ & 16 & 19.1 & 21.8 & 71.8 \\
    $\tau_\eta~[\text{yrs}]$ & 47.8 & 31.6 & 94.2 & 38.3 \\
    $\tau~[\text{yrs}]$ & 12 & 11.9 & 17.7 & 25 \\
    $\Delta\phi~[^\circ]$ & 159 & 164 & 158 & 128 \\
\hline                                         
\end{tabular}
\end{table}

Fig. \ref{fig:hath_lin} compares the observed rate of the equatorial drift of sunspot zones (Eq. (\ref{eq:hlaw})) with that of the toroidal flux system for the models discussed above. The model with $n=1$ matches the observed equatorial drift well, with models with other $n$ matching less well.

\begin{figure}
\resizebox{\hsize}{!}{\includegraphics{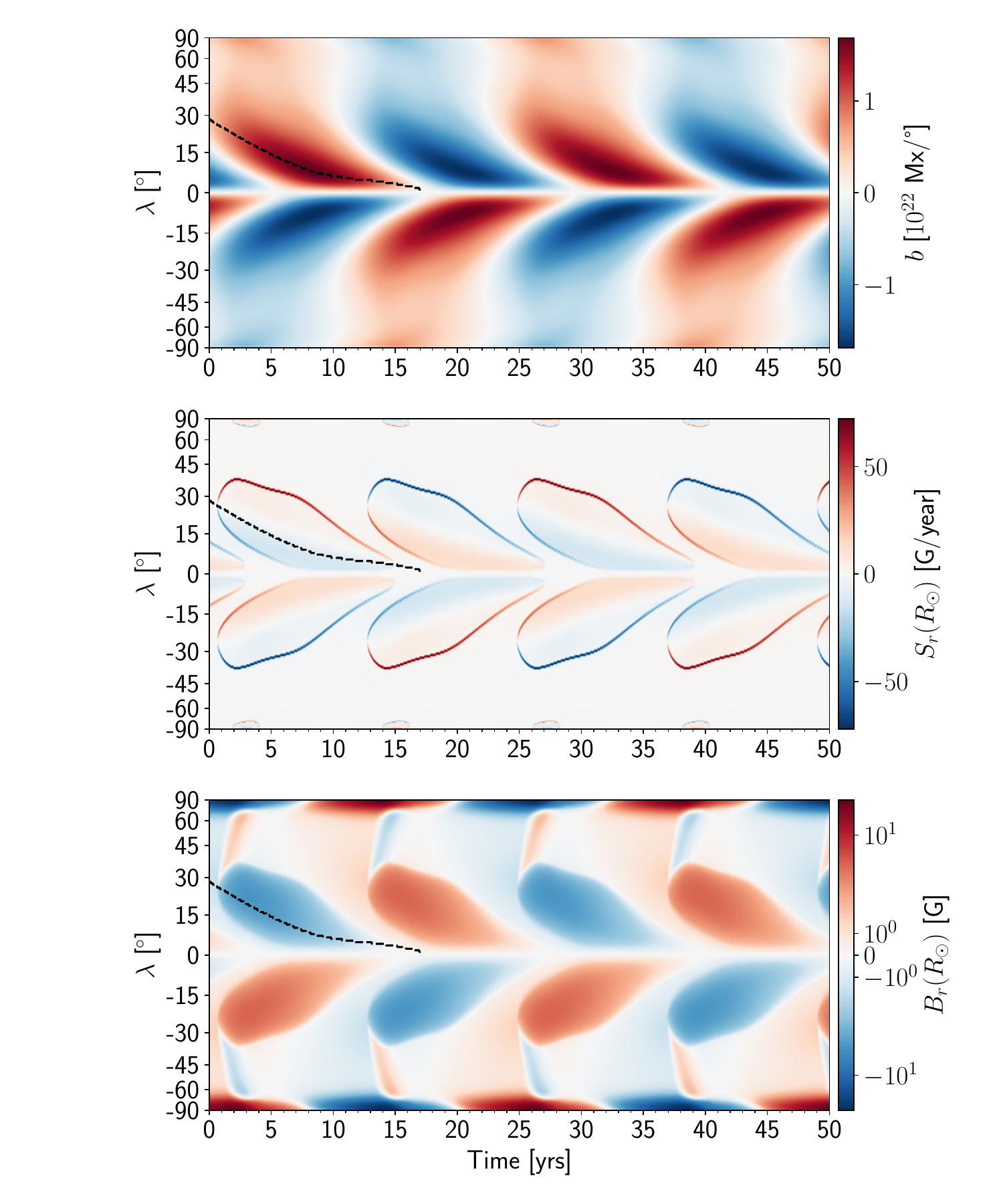}}
\caption{Time-latitude diagrams of the toroidal flux density $b$ (top), surface radial source term $S_r(R_{\odot})$ (middle), and surface radial field $B_r(R_{\odot})$ (bottom) for Model TA.}
\label{fig:t1}
\end{figure} 

\begin{figure}
\resizebox{\hsize}{!}{\includegraphics{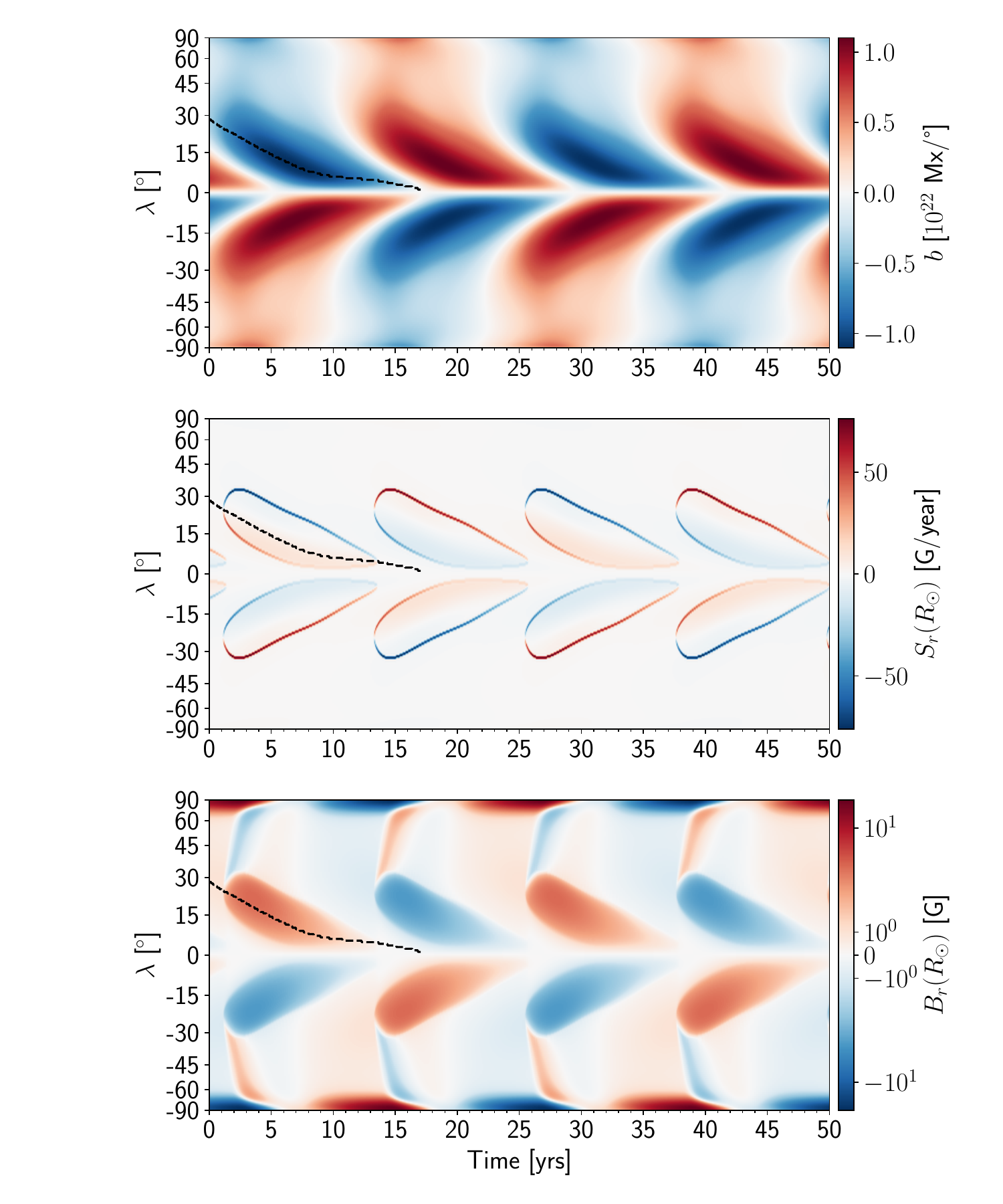}}
\caption{Time-latitude diagrams of the toroidal flux density $b$ (top), surface radial source term $S_r(R_{\odot})$ (middle), and surface radial field $B_r(R_{\odot})$ (bottom) for Model TB.}
\label{fig:t1m}
\end{figure} 

\begin{figure}
\resizebox{\hsize}{!}{\includegraphics{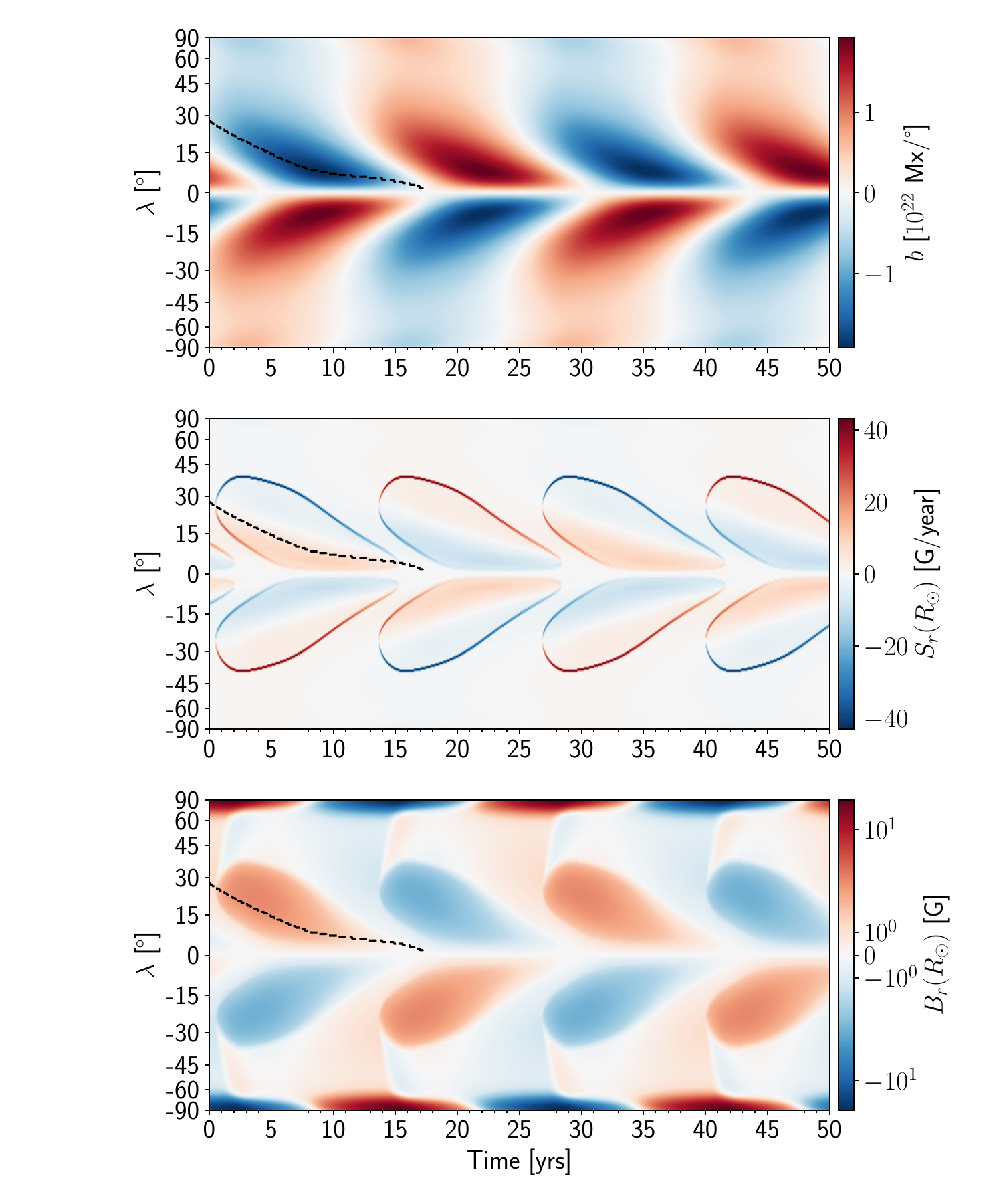}}
\caption{Time-latitude diagrams of the toroidal flux density $b$ (top), surface radial source term $S_r(R_{\odot})$ (middle), and surface radial field $B_r(R_{\odot})$ (bottom) for Model TC.}
\label{fig:d2}
\end{figure}

\begin{figure}
\resizebox{\hsize}{!}{\includegraphics{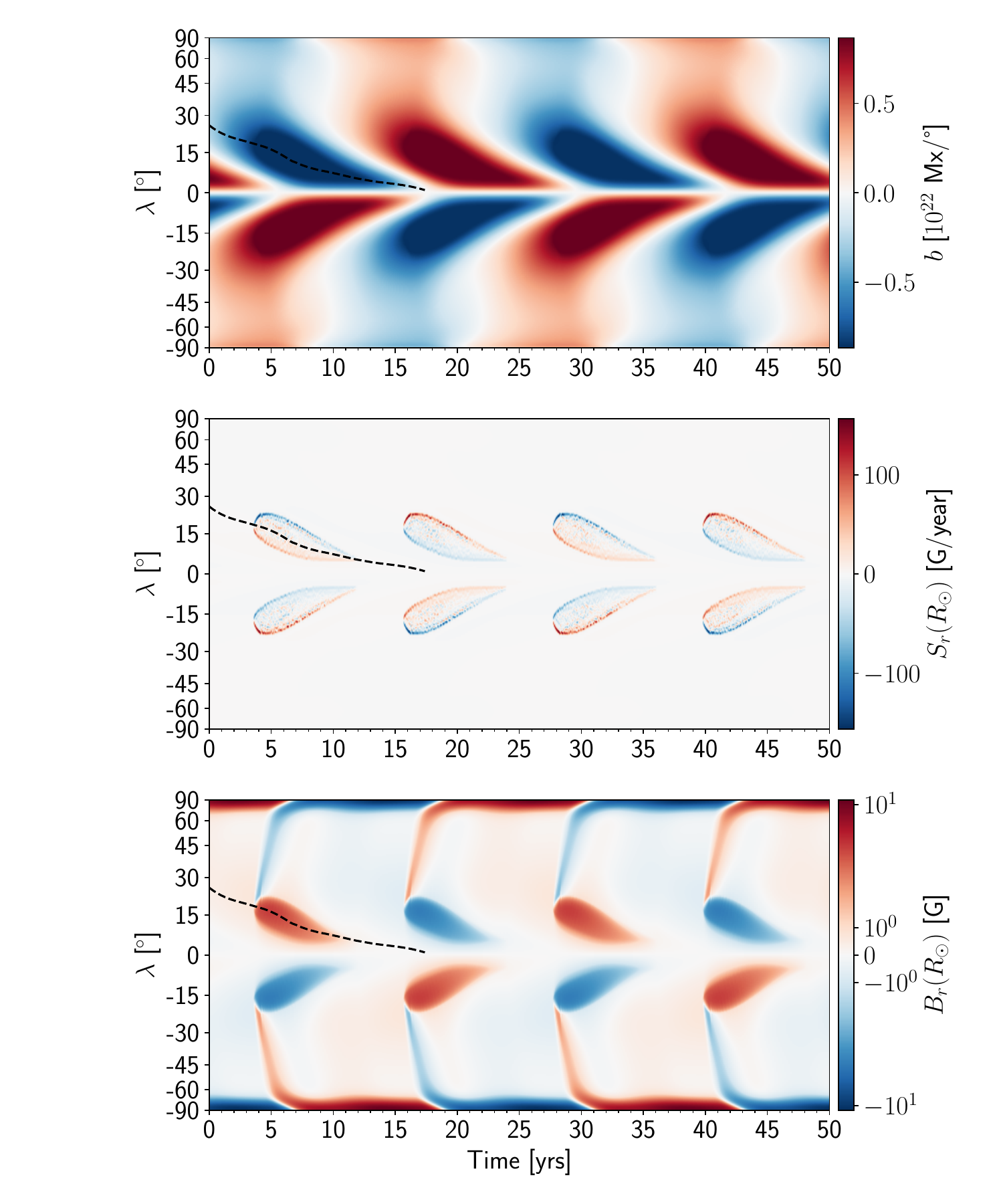}}
\caption{Time-latitude diagrams of the toroidal flux density $b$ (top), surface radial source term $S_r(R_{\odot})$ (middle), and surface radial field $B_r(R_{\odot})$ (bottom) for Model TD.}
\label{fig:t3}
\end{figure}

\subsection{Models with either slow or fast emergence based on the toroidal flux density} \label{sect:res_thresh}
In this section we show a selection of models incorporating the two-regime threshold described in Section \ref{sect:modtworeg}. We computed grids of models with different values of $\tau_0$ and $\gamma_0$ for two values of the threshold field, $B_{\text{thresh}}=1$~kG and $1.35$~kG, and bulk diffusivity, $\eta_{\text{CZ}}=10$ and 20 km$^2$/s. We present four of the models that best fit a number of observational constraints: the surface radial field inside the butterfly wings of $\sim 5~\text{G}$, polar fields that are not too strong, a cycle period reasonably close to 12 years, and  little flux emergence in the polar regions. The four solutions are presented in Figs. \ref{fig:t1}, \ref{fig:t1m}, \ref{fig:d2}, and \ref{fig:t3}, which we will respectively call Model TA, TB, TC, and TD. The parameter values are found in Table \ref{table:thresh}. The different output quantities are presented in Table \ref{table:thresh_out}.

All models now present a clear, strong, initial rush to the poles. This is followed by a weak poleward surge of leading polarity near activity maxima, followed by more trailing polar flux until the cycle ends.

The source term shows strong poloidal flux production at the edges where the threshold condition is met. This is a consequence of the fact that the threshold introduces a discontinuity in $S$, which is the source term in the equation for $A$. In the absence of diffusion, the resulting discontinuity in $S$ leads to a delta function in $S_r$. The presence of diffusion smooths these singular features.

\begin{figure}
\resizebox{\hsize}{!}{\includegraphics{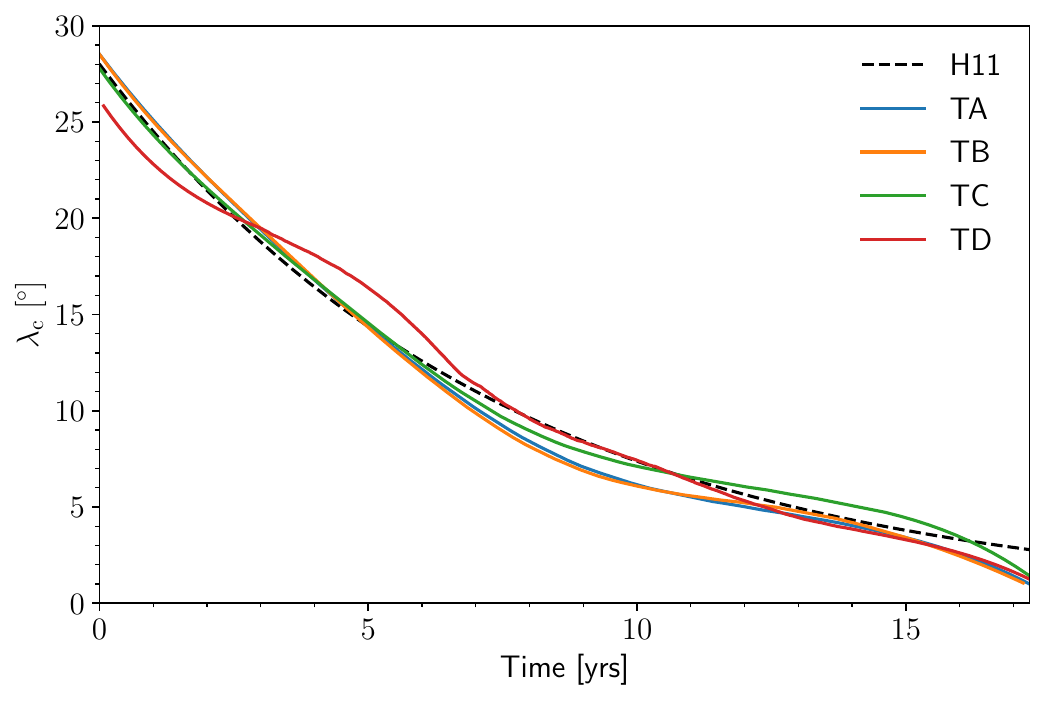}}
\caption{Comparison of the \citetalias{Hathaway2011} standard law of sunspot zone migration with that obtained by threshold models.}
\label{fig:hath_thresh}
\end{figure} 

As in \citet{Biswas2022}, the presence of a threshold in flux emergence and the flux depletion associated with it causes the dynamo to saturate. Because it is the amount of leading poloidal flux cancelling across the equator that determines the strength of the Sun's dipole, emergences at high latitudes are inefficient at generating polar fields \citep{Jiang2014}. As stronger cycles present emergences at higher latitudes than weaker ones, the polar field at cycle minimum is weaker and consequently so is the subsequent cycle. This saturation mechanism is known as latitudinal quenching \citep{Jiang2020,Karak2020,Talafha2022}. For the particular model parameters chosen here, latitudinal quenching, by itself, is not sufficient to saturate the dynamo. The addition of emergence loss causes the early, high-latitude emergences of stronger cycles to deplete the subsurface toroidal flux reservoir very quickly, enhancing the effect of latitudinal quenching. Note that the observational results of Waldmeier are seen as evidence for latitudinal quenching in the solar dynamo \citep{Waldmeier1955,CS2023}. As explained by \citet{Biswas2022}, the non-linearity involving emergence loss and the threshold makes the decline phase independent of cycle strength. 

We briefly explored the effects of varying the model parameters. Decreasing $\tau_0^\text{fast}$ causes a narrowing of the butterfly wings (compare Figs. \ref{fig:t1m} and \ref{fig:t3}), while increasing $\gamma_0$ causes not only a widening of the butterfly wings, but a stronger one during the ascending phase of the cycle. Increasing $\eta_\text{CZ}$ makes the period longer and necessitates a significant increase of pumping. $B_\text{thresh}$ essentially sets the normalization of the magnetic field. Models with values of $B_\text{thresh}$ much larger than 1 kG have fields much too strong compared to observations. This value of $B_\text{thresh}$ is significantly weaker than the equipartition value of $5-6$ kG as inferred from MLT (although it is in practice likely quenched by rotation and magnetic fields).

Fig. \ref{fig:hath_thresh} compares the \citetalias{Hathaway2011} law with the equatorward migration of the toroidal flux system of threshold models. All models reproduce the observed behaviour reasonably well.

\begin{table}
\caption{Input parameters of magnetic buoyancy models}
\label{table:buoy}
\centering
\begin{tabular}{c c c c c c c}
\hline\hline
Model & BA & BB & BC & BD & BE & BF \\
\hline       
    $\tau_0^b~[\text{yrs}]$ & 72 & 72 & 72 & 72 & 72 & 72 \\
    $\gamma_0~[\text{m/s}]$ & 30 & 30 & 30 & 30 & 100 & 30 \\
    $\eta_\text{CZ}~[\text{km}^2/\text{s}]$ & 35 & 35 & 10 & 10 & 35 & 35 \\
    $L$ & on & off & on & on & on & on \\
    $m$ & 2 & 2 & 2 & 1 & 2 & 2 \\
    cycle & $\langle\rangle$ & $\langle\rangle$ & $\langle\rangle$ & $\langle\rangle$ & $\langle\rangle$ & 23 \\
\hline                                          
\end{tabular}
\end{table}

\begin{table}
\caption{Output quantities of magnetic buoyancy models}
\label{table:buoy_res}
\centering
\begin{tabular}{c c c c c c c}
\hline\hline
Model & BA & BB & BC & BD & BE & BF \\
\hline
    $P~[\text{yrs}]$ & 14.6 & 9.2 & 10.7 & 10.6 & 13.3 & 12  \\
    $B_p~[\text{G}]$ & 14.2 & 37.3 & 37.3 & 88.5 & 17.8 & 25.7 \\
    $B_b~[\text{G}]$ & 2 & 17.6 & 12.3 & 28.4 & 2.9 & 4 \\
    $\Phi_\text{m}~[10^{23}~\text{Mx}]$ & 5.3 & 9.9 & 9.9 & 26.5 & 6.2 & 7 \\
    $\text{max}(R)$ & 104 & 622 & 489 & 1405 & 159 & 206 \\
    $\text{min}(R)$ & 9 & 165 & 111 & 515 & 16 & 21 \\
    $\Delta R$ & 0.09 & 0.27 & 0.23 & 0.37 & 0.10 & 0.10  \\
    $\tau_L~[\text{yrs}]$ & 31.9 & -- & 14 & 12.5 & 24.6 & 21.3 \\
    $\tau_\eta~[\text{yrs}]$ & 173.5 & 27 & 35.6 & 37.1 & 244 & 118 \\
    $\tau~[\text{yrs}]$ & 27 & 27 & 10.1 & 9.4 & 22.4 & 18.1 \\
    $\Delta\phi~[^\circ]$ & 138 & 144 & 155 & 170 & 146 & 157 \\
\hline
\end{tabular}
\end{table}

\begin{figure}
\resizebox{\hsize}{!}{\includegraphics{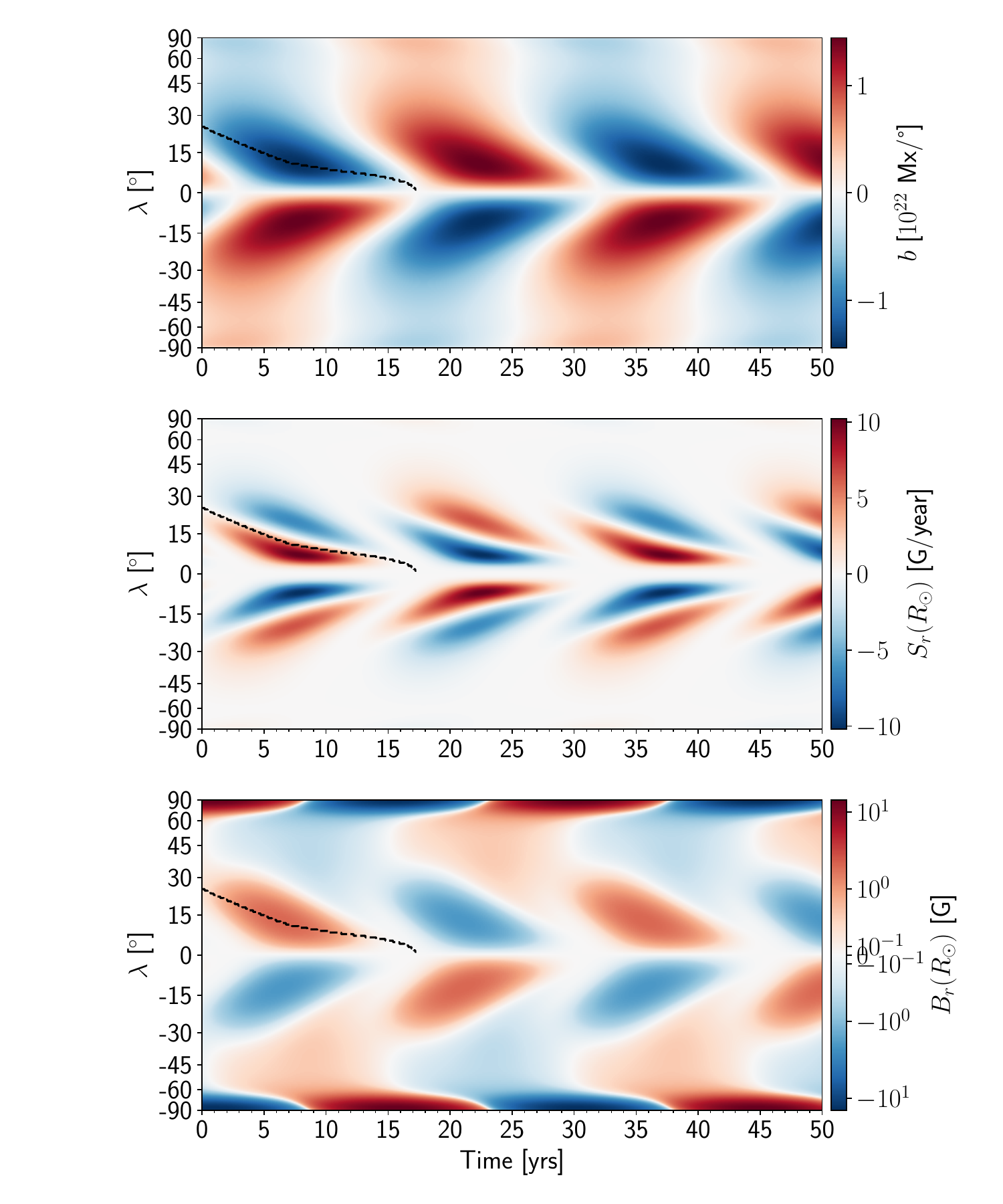}}
\caption{Time-latitude diagrams of the toroidal flux density $b$ (top), surface radial source term $S_r(R_{\odot})$ (middle), and surface radial field $B_r(R_{\odot})$ (bottom) for Model BA.}
\label{fig:buoy_34}
\end{figure} 

\begin{figure}
\resizebox{\hsize}{!}{\includegraphics{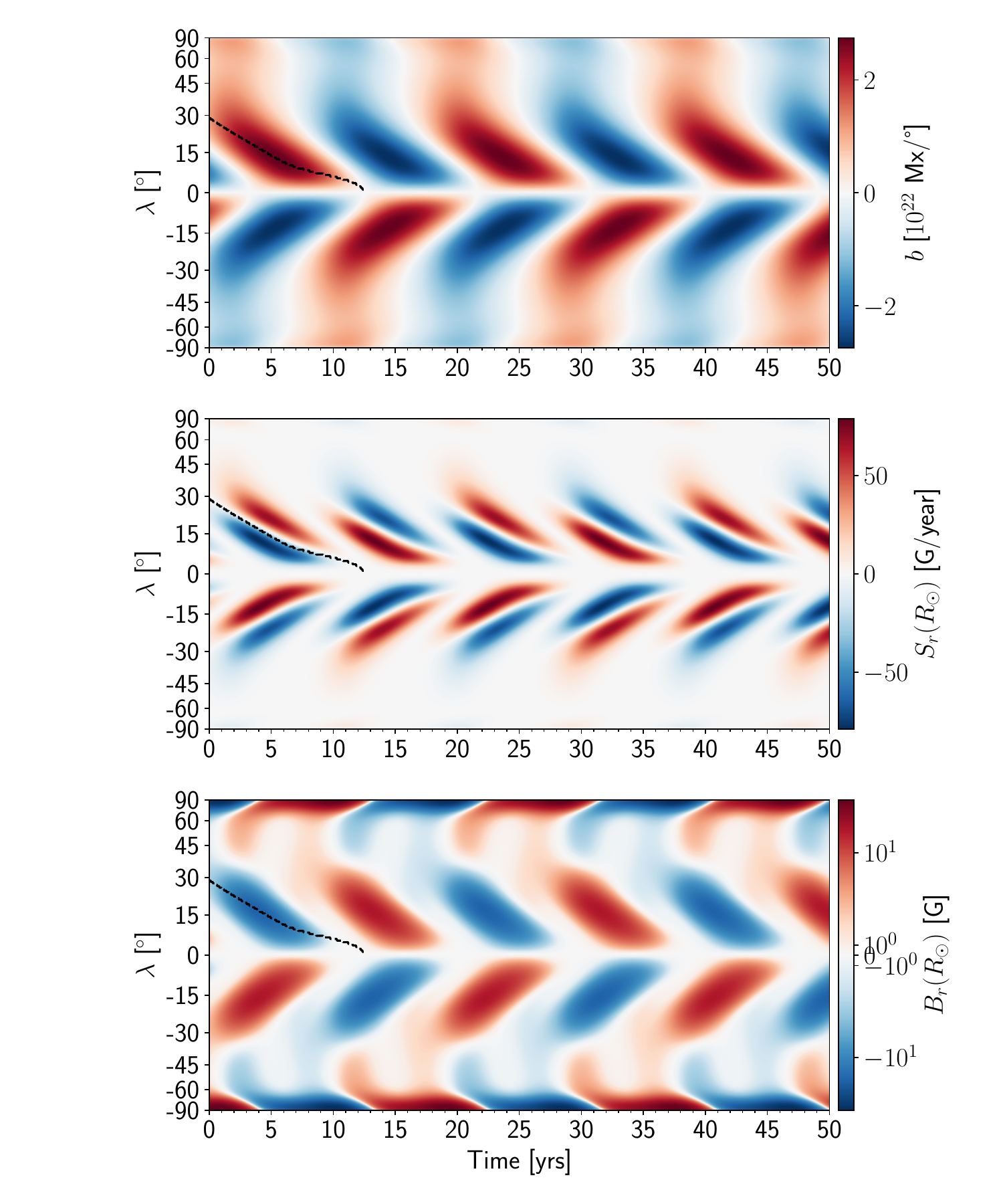}}
\caption{Time-latitude diagrams of the toroidal flux density $b$ (top), surface radial source term $S_r(R_{\odot})$ (middle), and surface radial field $B_r(R_{\odot})$ (bottom) for Model BB.}
\label{fig:buoy_34_l}
\end{figure} 

\begin{figure}
\resizebox{\hsize}{!}{\includegraphics{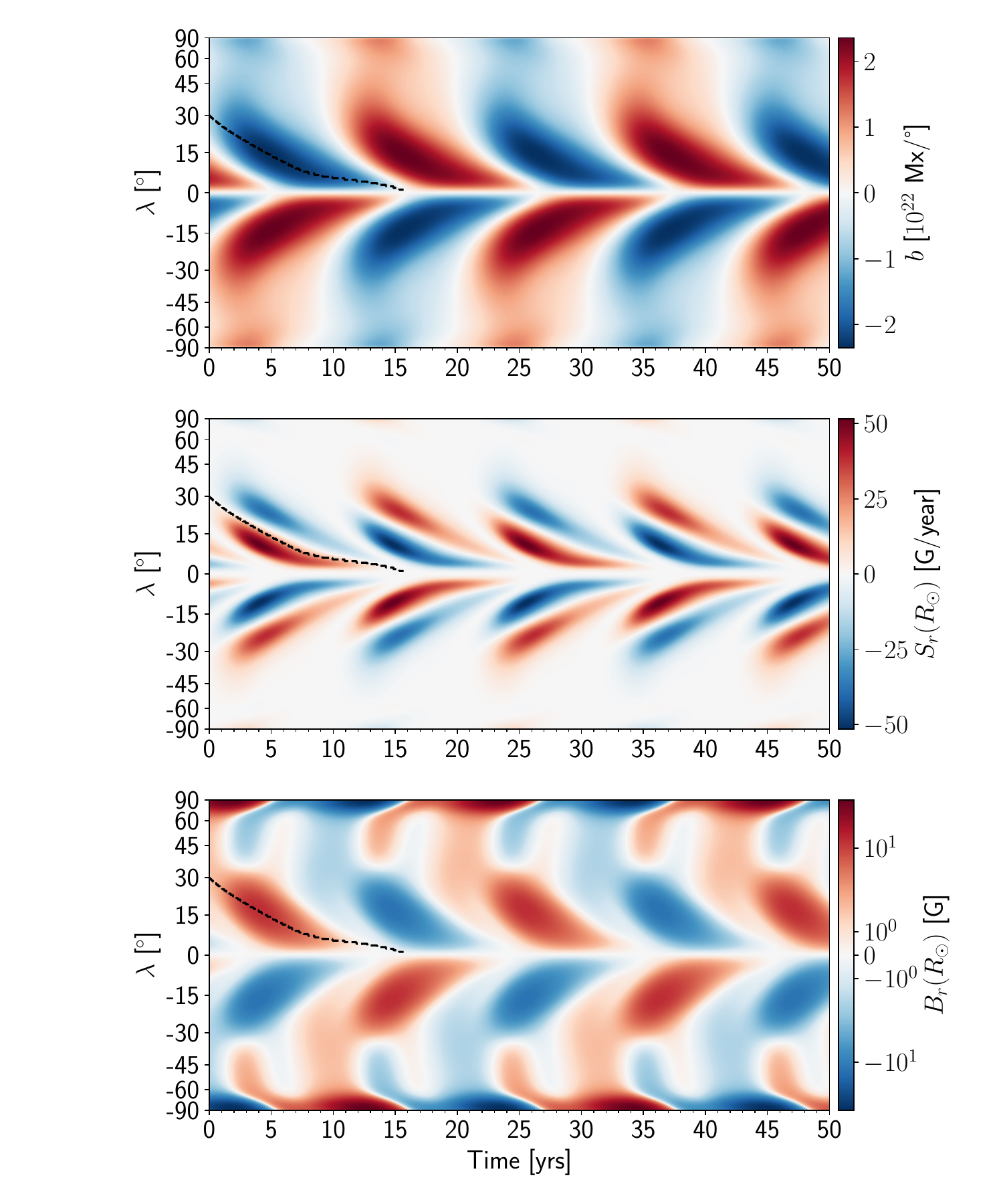}}
\caption{Time-latitude diagrams of the toroidal flux density $b$ (top), surface radial source term $S_r(R_{\odot})$ (middle), and surface radial field $B_r(R_{\odot})$ (bottom) for Model BC.}
\label{fig:buoy_34_e}
\end{figure} 

\begin{figure}
\resizebox{\hsize}{!}{\includegraphics{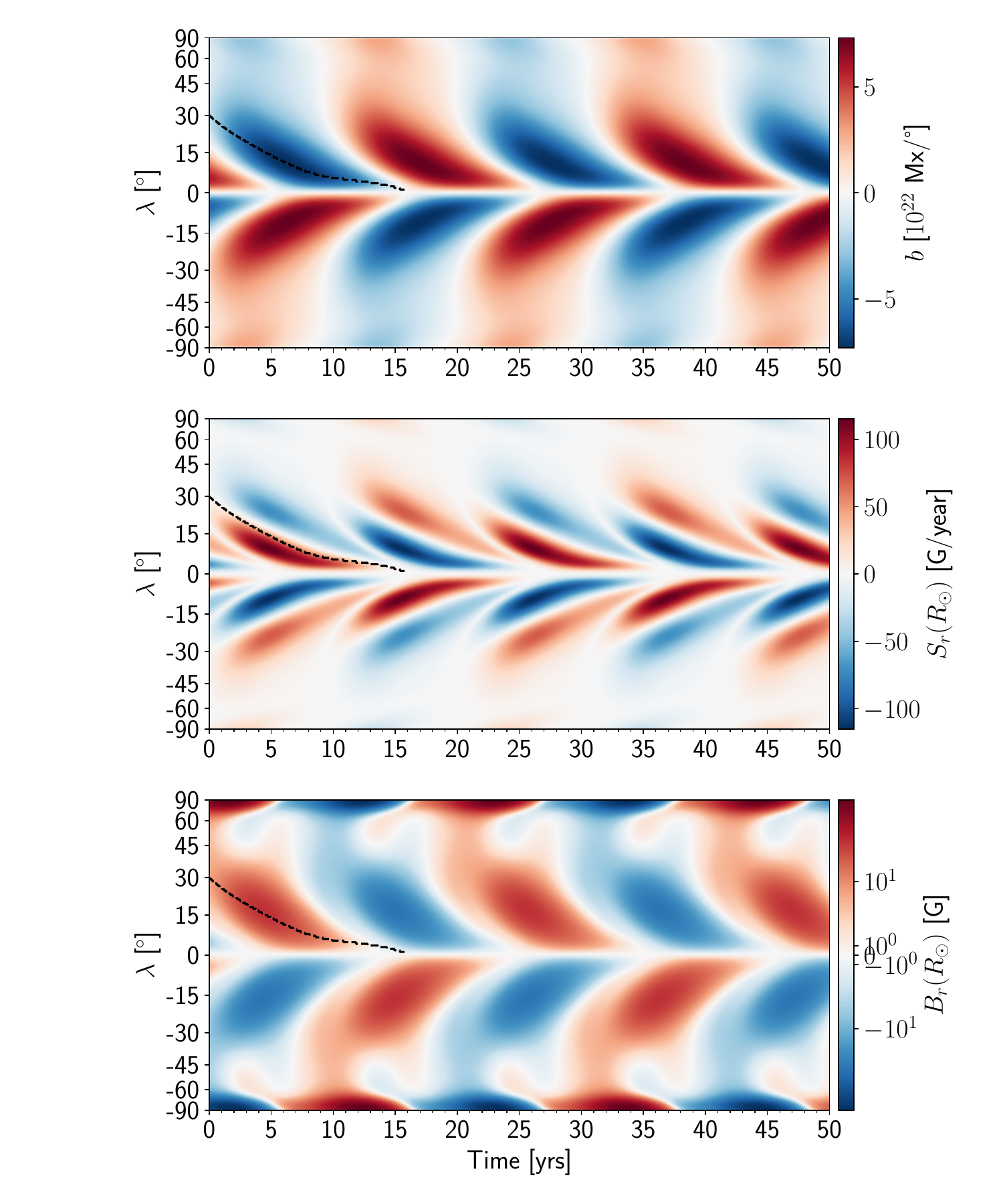}}
\caption{Time-latitude diagrams of the toroidal flux density $b$ (top), surface radial source term $S_r(R_{\odot})$ (middle), and surface radial field $B_r(R_{\odot})$ (bottom) for Model BD.}
\label{fig:buoy_34_m}
\end{figure} 

\subsection{Models where the emergence rate depends on magnetic buoyancy} \label{sect:res_buoy}

In this section we show a selection of models incorporating magnetic buoyancy in the source and loss terms, as described in Section \ref{sect:buoy}. Contrary to Section \ref{sect:res_thresh}, all the models we present have the same values of $\tau_0^b=72~\text{yrs}$ and $\gamma_0=30~\text{m/s}$. The model that best reproduce the observations with a "reasonable" pumping amplitude will be our "reference model", with $m=2$ and $\eta_\text{CZ}=35~\text{km}^2/\text{s}$ (model BA). We then change individually a few parameters to study their effect on the solution; the loss term switched off $L=0$ (model BB), a weaker diffusivity $\eta_\text{CZ}=10~\text{km}^2/\text{s}$ (model BC),  and the emergence rate proportional to the toroidal field strength $m=1$ (model BD -- here a lower diffusivity is necessary). They are presented in Figs. \ref{fig:buoy_34} to \ref{fig:buoy_34_m}. Two additional models, which do not present sufficiently different time-latitude diagrams to be shown, have been computed. They are Model BE having a pumping amplitude of $\gamma_0=100~\text{m/s}$ and Model BF using the meridional flow profile of cycle 23. At mid-latitudes at the bottom of the convection zone, the amplitude of the meridional flow is 4.8 m/s for cycle 23 and 3.6 m/s for cycle 24, meaning that in model BF this velocity is increased from their average by 0.6 m/s. The input parameters and output quantities are presented in Tables \ref{table:buoy} and \ref{table:buoy_res}, respectively.

\begin{figure}
\resizebox{\hsize}{!}{\includegraphics{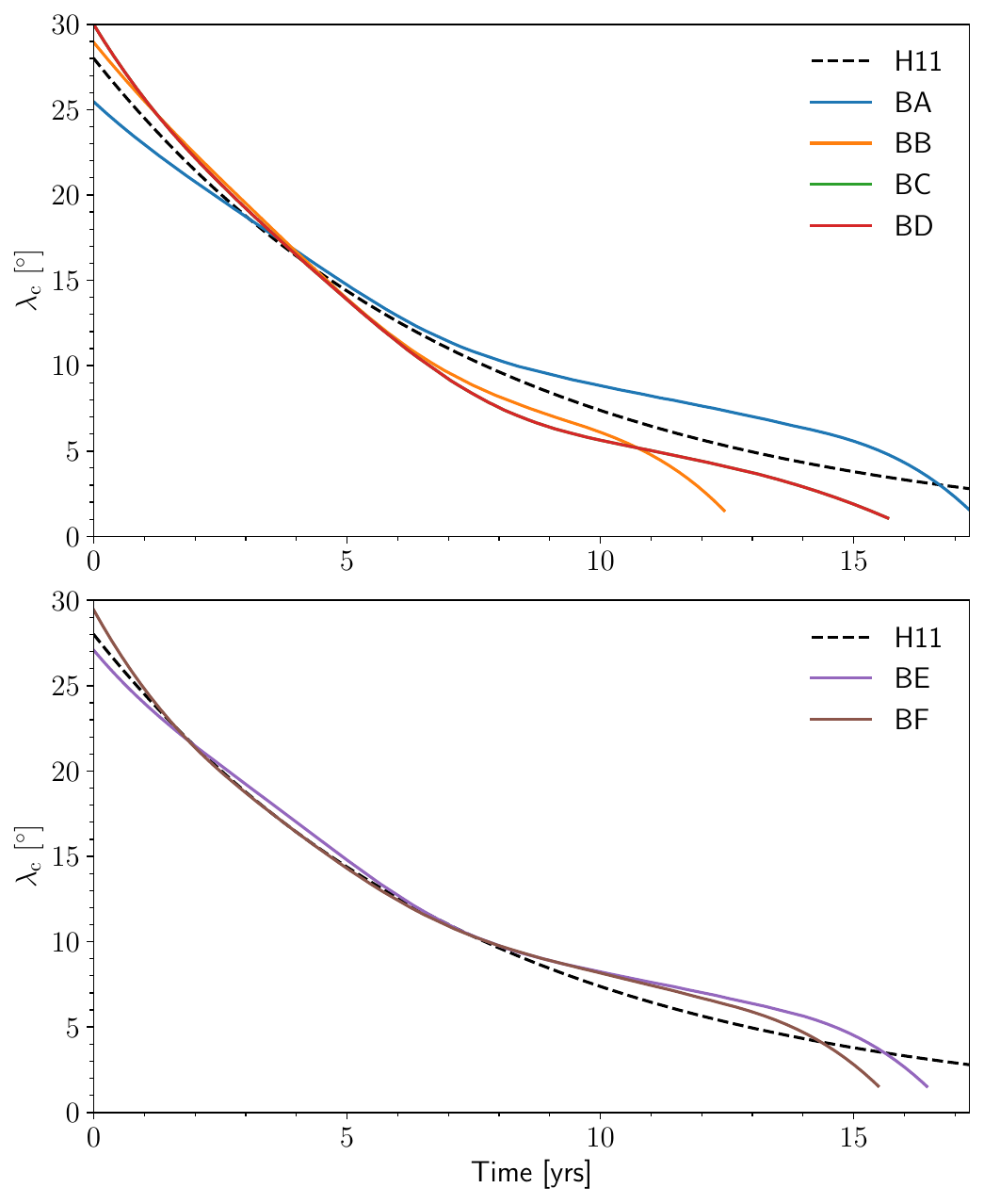}}
\caption{Comparison of the \citetalias{Hathaway2011} standard law of sunspot zone migration with that obtained by buoyancy models.}
\label{fig:hath_buoy}
\end{figure} 

Model BA is, out of all the models presented in this paper, the most solar-like. It reproduces a clear rush to the poles without any opposite polarity surge. The width of the butterfly wings is somewhat below $30^\circ$, which is in good agreement with what is found in \citetalias{Cloutier2024} for the observed mean butterfly diagram (see also Fig.~\ref{fig:sr_obs_hath}), and is only weakly dependent on the model parameters. With $B_b=2~\text{G}$, the field strength inside the butterfly wings is consistent with that of the observed mean butterfly diagram of \citetalias{Cloutier2024} (cf. Fig. \ref{fig:sr_obs_hath}). With $B_p=14.2~\text{G}$, the polar field strengths are also consistent with observations. The maximal value of the net toroidal flux in one hemisphere is $\text{max}(\Phi)=5.3\times 10^{23}~\text{Mx}$, which is in good agreement with the estimates of \citet{CS2015}. The phase difference between the poloidal and toroidal fields, however, remains large at $\Delta\phi=138^\circ$. This could be due to our models being close to symmetric with respect to cycle maximum. A faster rising phase would imply a stronger rush to the poles, reversing the polar field more quickly. The cycle period is also rather long at 14.6 yrs. It is possible to decrease the cycle period by either increasing turbulent pumping or decreasing the diffusivity. The latter makes the model less solar-like (see Fig. \ref{fig:buoy_34_e}), while the former requires excessively strong pumping velocities to significantly lower the cycle period. The cycle period is \textit{independent} of the value of $\tau_0^b$. This is because $\tau_0^b$ combines with $b_\text{eq}$, which acts to normalize the magnetic field strength. The value of $\tau_0^b$ was chosen so that $B_b=2~\text{G}$. The emergence loss timescale is also constant during the declining phases at $\tau_L=32.4~\text{yrs}$, where $\tau_\eta\simeq 180~\text{yrs}$ so that $\tau\simeq 27~\text{yrs}$. This is very close to the magnetic period of 29.2 years, and only a factor of two larger than the rough 12-year estimate of \citet{CS2020}.

To investigate the effect of the loss term on the solutions, we computed Model BB (Fig. \ref{fig:buoy_34_l}) where the loss term has been turned off. Unlike for the threshold prescription, the dynamo still saturates. A more precise look into the saturation of the buoyancy dynamo is presented in Appendix \ref{sect:app}. Switching off the loss term also leads to a shorter period, a full 5 years shorter (compare with \citetalias{Cloutier2023}). Emergence loss could therefore play an important role in setting the cycle period.

Model BC (Fig. \ref{fig:buoy_34_e}) has a lower value of the bulk diffusivity. The activity period is significantly decreased and is now only 10.7 years. Increasing the diffusivity, like turning on the loss term, spreads out the subsurface toroidal field, slowing down its build up near the equator. Weaker diffusivity also makes the emergence loss timescale $\tau_L$ much closer to the activity period (because the early strong emergences deplete the toroidal flux very quickly), and the total toroidal flux loss timescale $\tau$ slightly below it.

We also tested the effect of the value of $m$ for the non-linearity and computed Model BD (Fig. \ref{fig:buoy_34_m}) where $m=1$. The butterfly wings are slightly wider than those of Model BC because of the decreased non-linearity. Interestingly, all output quantities of Model BD in Table \ref{table:buoy_res} are very close to those of Model BC, which is the same save for a value of $m=2$, except for the quantities depending on the amplitude of the magnetic field. For those, they are a bit more than a factor of two larger, so that decreasing $\tau_0^b$ can bring them in relatively good agreement. Consequently, the degree of non-linearity has a weak effect on the solutions.

Increasing the turbulent pumping amplitude by more than a factor of two (Model BE) decreases the period by only 1.3 years. At a large value of 30 m/s, the time required for the magnetic field to reach the lower convection zone is already relatively close to instantaneous with respect to the cycle period. 

Using instead the meridional flow profile of cycle 23 (Model BF), the period is decreased by a full 2.5 years, illustrating the key role of the meridional flow on setting the cycle period. A faster meridional flow also significantly decreases the emergence loss timescale compared to the activity period. This is because by compressing the toroidal field closer to the equator, a faster meridional flow allows for stronger emergences to deplete the subsurface toroidal field more quickly. We comment here that the flow amplitudes at the bottom of the convection zone for cycles 23 and 24 are both within the error bars of the helioseismic inversions of \citet{Gizon2020}, despite the resulting periods differing by 2.5 years. 

Comparing Figs. \ref{fig:buoy_34} and \ref{fig:sr_obs_hath}, we see that the agreement between our buoyancy model and observations is remarkable. The shape of the butterfly wings and their width is very well reproduced. But the model surface radial source term is also very similar to the observed one (obtained by "inverting" a 1D surface flux transport model with the observed butterfly diagram -- middle panel of Fig. \ref{fig:sr_obs_hath}). Even the location of maximum toroidal field (corrected for the low resolution) is in agreement. 

In Fig. \ref{fig:hath_buoy} we compare the equatorial migration of the toroidal flux of our buoyancy models with the observed sunspot belt migration. For models BA to BD, the equatorial drift in the model is substantially faster than what is observed (consistent with the cycles being shorter). The match with observations is improved if the pumping is strongly increased to 100 m/s (Model BE). For Model BF with a faster meridional flow, the agreement is even better and remarkably good. The slowdown of the drift of the sunspot zones compared to Eq. (\ref{eq:hlaw}) about midway into the declining phase seen for models BA, BE and BF is a feature also appearing in the observations (cf. Fig. 5 of \citealt{Hathaway2011} and Fig. 2 of \citetalias{Cloutier2024}).

The bulk turbulent diffusivity of $35~\mathrm{km}^2/\mathrm{s}$ used in our buoyancy models is about a factor of 30 lower than MLT estimates, $\sim 1000~\mathrm{km}^2/\mathrm{s}$ \citep[e.g.][]{MJ2011}. In principle, magnetic diffusivity can be significantly quenched both by rotation and the magnetic field \citep[][see \citealt{Cloutier20230} for a discussion on the matter]{Kitchatinov1994,FH2016,HK2021}. Furthermore, the corresponding MLT convective length scale is that of the giant cells, which seem to be ruled out by observations \citep{Hanasoge2012,Gizon2021}. Mean-field theory predicts turbulent diffusivity to be given by $\eta=\frac{1}{3}\overline{v_c^2}\tau_c$ \citep[e.g.][]{Kitchatinov1994}, where $\tau_c$ is the convective correlation time. Its value can be estimated from helioseismology or local correlation tracking. With values of $v_c\simeq 10~\mathrm{m/s}$ and $\tau_c\simeq 1$ month \citep{HU2021}, $\eta\simeq 86~\mathrm{km}^2/\mathrm{s}$. This is around a factor of 2 larger than the value used in our buoyancy models, but is a factor of 10 smaller than the MLT estimate. Lastly, we mention again the possible role of the helioseismically-inferred meridional flow in contributing to most of the total diffusivity in our models (Sect. \ref{sect:turb}).

\section{Conclusion}
The motivation behind this paper was to identify mechanisms by which the rush to the poles could be reproduced in BL-FTD models. A mechanism either suppressing at high latitudes or enhancing at low latitudes emergences is necessary, even if the toroidal field is weak close to the poles and mostly stored near the equator. An emergence probability quickly decreasing with latitude is one such mechanism. The physical motivation in this case would be related to the latitudinal dependence of the growth rate of the instability giving rise to the emergence of flux tubes. Alternatively, non-linear mechanisms such as a threshold in flux emergence and an emergence rate based on magnetic buoyancy, not only help reproduce the rush to the poles, but provide a saturation mechanism for the dynamo. They produce latitudinal quenching. Latitudinal quenching is believed to be related to the solar cycle property that all cycles decline the same way \citep{CS2023}. 

The "best fit" model butterfly diagrams making use of the buoyancy prescription (Fig. \ref{fig:buoy_34}) is strikingly similar to the mean observed one (Fig. \ref{fig:sr_obs_hath}), as are the poloidal field generation rates. The width of the butterfly wings is found to be only weakly dependent on model parameters and is around $\lesssim\pm 30^\circ$, in good agreement with observations. It is the saturation of the dynamo that controls this width. Moreover, the equatorward drift of the activity belts is found to be in good agreement with that inferred from observations, implying the toroidal field could be stored at equatorial latitudes deep in the convection zone.

An interesting finding of this paper is just how much the depletion of toroidal flux deep in the convection zone leading to the emergence of poloidal flux at the surface lengthens the cycle period for non-linear models. It may be slowed down by as much as $\simeq60\%$, with the timescale associated with the toroidal flux loss through the surface being comparable to the magnetic period. What sets the cycle period could hence diverge from the usual dynamo-wave-FTD dichotomy, emergence loss possibly playing an important role.

We stress that our model incorporates the observed axisymmetric flows (differential rotation and meridional circulation). For this reason, our model cannot be easily used to draw inferences about the dynamos of other stars. At the very least an extension which models the differential rotation and meridional circulation of other stars would be necessary.

\begin{acknowledgements}
The authors wish to thank the anonymous referee for comments that helped improve the overall quality of this paper. This work was carried out when SC was a member of the International Max Planck Research School for Solar System Science at the University of Göttingen. The authors acknowledge partial support from ERC Synergy grant WHOLE SUN 810218.
\end{acknowledgements}

\bibliographystyle{aa}
\bibliography{Cloutier_2024b}

\begin{appendix}
\section{Saturation of the buoyancy dynamo} \label{sect:app}

\begin{figure}
\resizebox{\hsize}{!}{\includegraphics{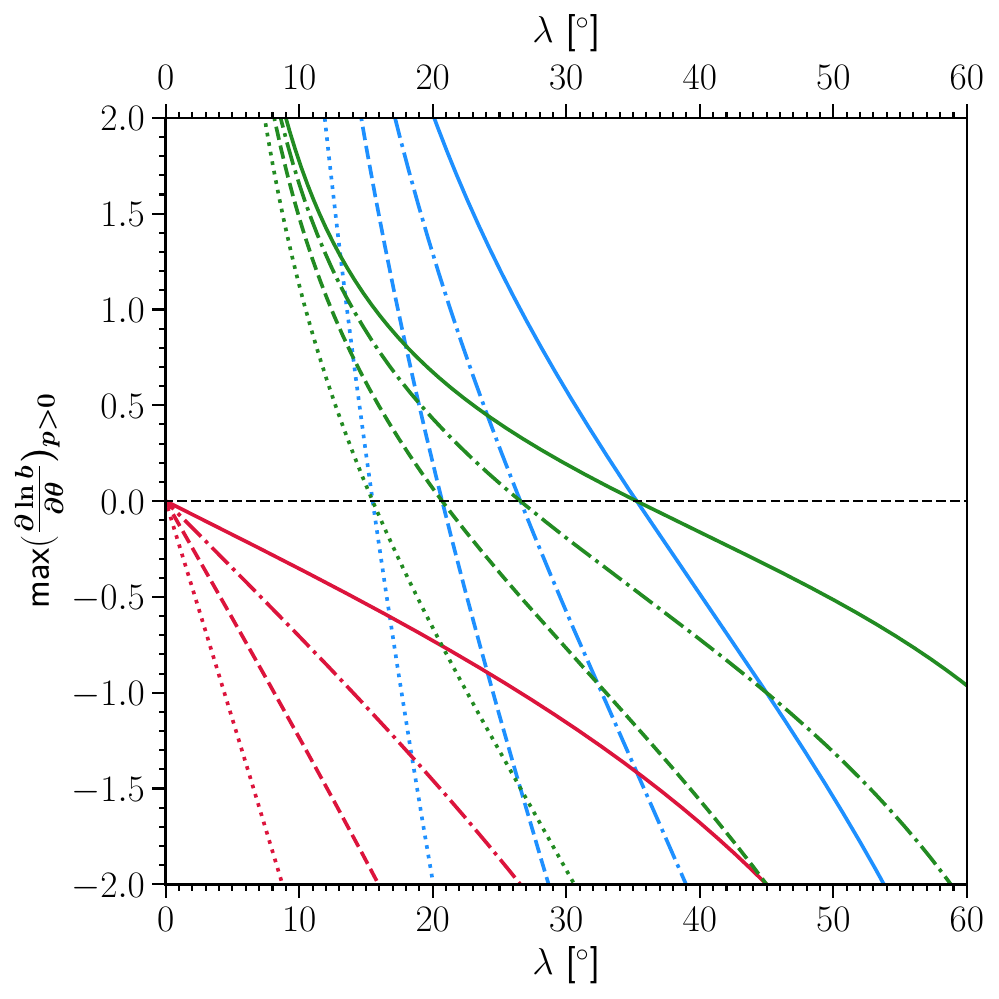}}
\caption{Maximal value of $\partial\ln b/\partial\theta$ possible for the generation of leading polarity field at latitude $\lambda$ for values of $n=1,3,6,12$ (solid, dash-dotted, dashed and dotted lines respectively). Blue, green and red lines represent cases where $m=0$, $m=2$ and constant tilt angle.}
\label{fig:p}
\end{figure} 

That our models with the buoyancy prescription ($m>0$) saturate even with the loss term turned off, means that the latitudinal quenching of the buoyancy dynamo is much stronger than that of the threshold dynamo. To shed light on this matter, it is advantageous to look not into the usual source term for the $\phi$-component of the poloidal vector potential $A$ (Eq. (\ref{eq:s})), but into the source term for the radial field at the surface (Eq. (\ref{eq:sr})). It can be rewritten as
\begin{equation} \label{eq:sra}
\begin{aligned}
    S_r(R_{\odot},\theta,t)=&\frac{1}{R_{\odot}^2\sin\theta}\frac{\partial}{\partial\theta}\left(\sin^{n+1}\theta\sin\delta\left|\frac{b(\theta,t)}{b_{\text{eq}}}\right|^m\frac{b(\theta,t)}{\tau_0^b}\right)\\
    =&-p(\theta,t)\sin^n\theta\sin\delta\left|\frac{b(\theta,t)}{b_{\text{eq}}}\right|^m\frac{b(\theta,t)/R_{\odot}^2}{\tau_0^b},
\end{aligned}
\end{equation}
where
\begin{equation} \label{eq:pfunc}
    p(\theta,t)=-\left((n+1)\cot\theta+\frac{\partial\ln(\sin\delta)}{\partial\theta}+(m+1)\frac{\partial\ln b}{\partial\theta}\right).
\end{equation}
$S_r$ can be rewritten in a way formally similar to $S$ (second equality) by introducing the polarity function $p$, which determines the polarity of the radial field being generated. The minus sign was chosen so that for $p(\theta,t)>0$ the polarity being generated is that of the leading spot (in the Northern hemisphere). 

Setting $p=0$ yields the latitude where the switch in polarity of the average BMR occurs, in other words the location of the center of the mean BMR being generated. We thus have a condition on the maximal value of $\partial\ln b/\partial\theta$ for leading spot polarity to be generated:
\begin{equation} \label{eq:p}
    \frac{\partial\ln b}{\partial\theta}<\frac{\tan\theta-(n+1)\cot\theta}{m+1},
\end{equation}
where $\sin\delta=\frac{1}{2}\cos\theta$ \citep{Leighton1969}. This condition is shown in Fig. \ref{fig:p} (blue and green lines). As we go equatorwards from the poles, $\partial\ln b/\partial\theta$ is positive and maximal close to the edge of the low-latitude toroidal flux concentration, zero where the latter is maximum, and negative very close to the equator. Wherever the maximal $b$-gradient is negative, trailing polarity is generated. The latitude where this gradient is zero, i.e. where the toroidal flux density $b$ is maximum, constitutes a lower bound on the latitude of the mean BMR center.

Not unexpectedly, forcing emergences to occur at lower latitudes by increasing the value of $n$ shifts this mean BMR towards the equator. One of the consequences is to increase the polar field strength, as cross-equatorward flux cancellation of the leading spot fields is increased. Increasing the non-linearity in $b$ of the source term, i.e. increasing the value of $m$, decreases the maximal $b$-gradient so that the center of the mean BMR is also shifted equatorward.

If Joy's law were to be constant with latitude (Fig. \ref{fig:p}, red lines), leading spot polarity would be produced only very close to the equator, after the maximum of $b$. If the average leading spot was to be produced so close to the equator, most of the leading flux would diffusively cancel across the equator and the resulting polar field would be extremely large, not to mention that the butterfly diagram would look very different from the observations. The exact form of Joy's law thus largely sets what the butterfly diagram looks like.

Examining Eqs. (\ref{eq:sra}) and (\ref{eq:pfunc}), we see that increasing the toroidal flux density $b$ does not shift the center of the mean BMR. Taking the derivative of the second form of Eq. (\ref{eq:sra}) with respect to $b$ yields
\begin{equation}
    \frac{\partial S_r}{\partial b}(R_\odot,\theta,t)\propto b^m(\theta,t).
\end{equation}
Increasing $b$ everywhere by the same factor causes a linear or non-linear increase of the radial source term (rather than a uniform increase), in such a way that the largest increase occurs near the mean BMR center. The field is thus redistributed closer to the center so that more intra-hemisphere flux cancellation occurs and the polar field is weakened; this constitutes a saturation mechanism. There is here an analogy with tilt quenching. It is as if the tilt of the mean BMR was decreased, leading to less cross-equator flux cancellation. This is, of course, not true tilt quenching, as the tilt of the individual emergences is constant in our models. The emergence rate simply changes as a function of latitude in such a way that the mean BMR's tilt appears smaller.  

But this is not the entire picture. While a proportional increase of $b$ does not change the location of the mean BMR being generated, the non-linearity of the BL source term leaves an imprint on the $\Omega$-effect and the toroidal field it induces for the next cycle. The distribution of the subsurface toroidal field changes as well. If, because of this non-linearity, significant toroidal field strengths are reached at higher latitudes, meaning that we have a stronger cycle with broader sunspot zones, the $b$-gradient is flattened, causing the mean BMR center to shift polewards. This, in turn, weakens the polar fields through increased cross-hemispheric flux cancellation of the generated radial field; this corresponds to latitudinal quenching (not only in the sense of the mean BMR). There is thus an interaction between the "tilt" and latitudinal quenchings. They, however, act against each other \citep[as in][where there is true tilt quenching]{KM2017}. The competition between both saturation mechanisms is what controls the width of the butterfly wings \citep[on that see also][although his model does not incorporate tilt quenching and the latitudinal quenching arises from a latitudinal dependence on the critical field strength required for flux emergence]{Karak2020}.
\end{appendix}

\end{document}